\documentclass[aps,prb,preprint,groupedaddress,citeautoscript,nopacs]{revtex4}

\usepackage{graphicx}
\usepackage{multirow}
\usepackage{color}
\usepackage{bm}
\usepackage{times}
\usepackage{amsmath,bm,amsfonts}
\usepackage{dcolumn}
\usepackage{graphicx}
\usepackage{latexsym}

\usepackage{ulem} 

\begin{document}

\title{Topological semimetals protected by off-centered symmetries in nonsymmorphic crystals}

\author{Bohm-Jung Yang$^{1,2,3,*}$}

\author{Troels Arnfred Bojesen$^{4,*}$}

\author{Takahiro Morimoto$^{5}$}

\author{Akira Furusaki$^{4,6}$}

\affiliation{$^1$ Department of Physics and Astronomy, Seoul National University, Seoul 08826, Korea}
\affiliation{$^2$ Center for Correlated Electron Systems, Institute for Basic Science (IBS), Seoul 08826, Korea}
\affiliation{$^3$ Center for Theoretical Physics (CTP), Seoul National University, Seoul 08826, Korea}
\affiliation{$^4$ RIKEN Center for Emergent Matter Science, Wako, Saitama, 351-0198, Japan}
\affiliation{$^5$ Department of Physics, University of California, Berkeley, CA 94720, USA}
\affiliation{$^6$ Condensed Matter Theory Laboratory, RIKEN, Wako, Saitama, 351-0198, Japan}
\affiliation{$^*$ These authors contributed equally to this work.}

\date{\today}

\begin{abstract}
Topological semimetals have energy bands near the Fermi energy
sticking together at isolated points/lines/planes
in the momentum space, which are often accompanied by 
stable surface states and intriguing bulk topological responses.
Although it has been known that certain crystalline symmetries
play an important role in protecting band degeneracy, 
a general recipe for stabilizing the degeneracy,
especially in the presence of spin-orbit coupling, is still lacking.
Here we show that a class of novel topological semimetals with point/line nodes
can emerge in the presence of {\it an off-centered rotation/mirror symmetry}
whose symmetry line/plane is displaced from the center of other symmorphic
symmetries in nonsymmorphic crystals.
Due to the partial translation perpendicular to the rotation
axis/mirror plane, an off-centered rotation/mirror symmetry
always forces two energy bands to stick together and form a doublet pair
in the relevant invariant line/plane in momentum space.
Such a doublet pair provides a basic building block for emerging
topological semimetals with point/line nodes
in systems with strong spin-orbit coupling.
\end{abstract}


\maketitle

Dirac particles with a pseudo-relativistic energy dispersion
has come to the fore in condensed matter physic research after the discovery of graphene~\cite{Graphene}.
To protect the four-fold degeneracy at a Dirac point in graphene, two conditions should be satisfied.
One is the simultaneous presence of time-reversal ($T$) and inversion ($P$) symmetries,
and the other is the absence of spin-orbit coupling.
When these two conditions are satisfied at the same time, the Berry phase around a Dirac point
has a quantized value of $\pi$, which guarantees the stability of the Dirac point.

Recently, there have been extensive efforts to extend
the physics of the two-dimensional (2D) graphene to three-dimensional (3D) systems~\cite{linenode_Kim,linenode_Weng,linenode_Yu,linenode_Schnyder,linenode_Xie,
linenode_Zheng,linenode_HY,linenode_Bian1,linenode_Bian2}.
A natural starting point is to search for a 3D Dirac point protected
by the $PT$ symmetry and the associated $\pi$ Berry phase.
Interestingly, however, it is found that the $PT$ symmetry protects a Dirac line node, instead of a Dirac point,
which gives rise to a 3D semimetal with Dirac line nodes
where four-fold band degeneracy occurs along a line in momentum space~\cite{linenode_Kim,linenode_Weng,linenode_Yu}.
As in the case of graphene, such a Dirac line node protected by the $PT$ symmetry is unstable in the presence of the spin-orbit coupling.
It is also reported that a Dirac line node can exist in systems with a mirror symmetry
when two bands with different mirror eigenvalues cross in the mirror plane~\cite{linenode_Schnyder,linenode_Xie,linenode_Zheng}.
However, the resulting line node is also unstable once the spin-orbit coupling is turned on.

In fact, the existence of a 3D Dirac point in systems with the spin-orbit coupling
requires the introduction of additional crystalline symmetries other than the time-reversal and the inversion symmetries~\cite{theory_Young,theory_Young2,LDA_Wang1,
LDA_Wang2,Yang_classification1,Yang_classification2}.
Up to now, two different recipes are known to yield 3D Dirac semimetals with point nodes.
One is to introduce an additional uniaxial rotation symmetry
where 3D Dirac points can occur when two bands with different rotation eigenvalues
cross on the rotation axis~\cite{LDA_Wang1,LDA_Wang2,Yang_classification1}.
Cd$_3$As$_2$ and Na$_3$Bi belong to this class~\cite{Na3Bi_Shen,Na3Bi_Hasan,Cd3As2_Hasan,Cd3As2_Cava,Cd3As2_Yazdani,Cd3As2_Chen}.
Although a Dirac point does not carry a nonzero monopole charge
which protects a Weyl point in the case of Weyl semimetals,
the rotation symmetry provides an integer topological charge at the Dirac point,
thus guarantees its stability~\cite{Yang_classification2}.

The second recipe is to introduce an additional
nonsymmorphic symmetry
such as glide mirrors or screw rotations.
When the double point group of a crystal possesses a four dimensional irreducible
representation, a Dirac point can appear at the Brillouin zone (BZ) boundary~\cite{theory_Young,theory_Young2}.
For several representative space groups,
projective symmetry group analyses have been performed,
which suggests $\beta$-BiO$_2$~\cite{theory_Young} and distorted spinel compounds~\cite{theory_Young2} as candidate systems
of 3D Dirac semimetals belonging to this class.
Since each Dirac point is protected by different combination
of crystalline symmetries depending on the space group of the crystal,
careful symmetry analysis is required, case by case, to find the relevant topological
charge of each Dirac point.

In this paper, we propose an alternative mechanism to realize
novel 3D semimetals with Dirac point/line nodes in systems with strong spin-orbit coupling
as well as $P$ and $T$ symmetries.
To protect nodal points/lines with four-fold degeneracy,
we find that {\it off-centered crystalline symmetries} play a crucial role.
In contrast to the case of ordinary glide mirror or screw rotation symmetries having
a partial translation in the invariant space of the associated point group symmetry,
an {\it off-centered rotation/mirror} symmetry involves a partial translation
that is orthogonal to the invariant space.
In centrosymmetric crystals, such an off-centered symmetry naturally arises
as a combination of a screw/glide symmetry and inversion symmetry $P$.
An off-centered mirror/rotation symmetry possesses
the characteristics of both the symmorphic and nonsymmorphic symmetries.
Namely, it has momentum independent quantized eigenvalues,
whereas its commutation relation with inversion symmetry $P$ depends on the momentum.
Due to such a mixed nature of the off-centered symmetry,
a pair of bands, each with Kramers degeneracy, form
{\it a doublet pair} in its invariant space in the first BZ,
and provide a basic building block for nodal points/lines.
Similarly, when the rotation axis (mirror plane) of a screw (glide) symmetry
does not pass the inversion center, an {\it off-centered screw (glide)} symmetry can be defined, which also leads to doublet pair formation
and emerging Dirac points (lines) in the relevant invariant space.  
When an external magnetic field is applied to these semimetals,
a Dirac-type point/line node with four-fold degeneracy
splits into two Weyl-type point/line nodes with two-fold degeneracy,
with emergent surface states connecting the split nodes.

\subsection*{Results}
{\bf Nature of off-centered rotation/mirror symmetries.}

Generally, a nonsymmorphic symmetry element $\widetilde{g}=\left\{g| \bm{t} \right\}$ is composed
of a point group symmetry operation $g$ and a partial lattice translation $\bm{t}=\bm{t}_{\perp}+\bm{t}_{\parallel}$
where $\bm{t}_{\parallel}$ ($\bm{t}_{\perp}$) is the component
invariant (variant) under the point symmetry operation $g$~\cite{book}.
For instance, in the case of a nonsymmorphic mirror symmetry $\widetilde{M}=\left\{M| \bm{t} \right\}$,
we have
\begin{align}
M\bm{t}_{\parallel}=\bm{t}_{\parallel}, \quad M\bm{t}_{\perp}=-\bm{t}_{\perp}.
\end{align}
Since $M^{2}=-1$ ($M^{2}=+1$) for particles with a half-integer (integer) spin,
when the nonsymmorphic mirror symmetry $\widetilde{M}=\left\{M| \bm{t} \right\}$ is operated
twice, it should be an element of the lattice translation group, i.e.,
$\left\{M| \bm{t} \right\}^{2}=\left\{M^{2}| 2\bm{t}_{\parallel} \right\}\in \mathbb{T}$
where $\mathbb{T}$ is the group of the pure lattice translation of a given crystal.
Thus $2\bm{t}_{\parallel}$ should be a unit lattice translation in the mirror invariant plane
whereas $\bm{t}_{\perp}$ is not influenced by the constraint above.

In fact, $\bm{t}_{\perp}$ is a fragile quantity whose value depends on the choice of
the reference point of the point group symmetry operation.
For instance, if the reference point for the point group symmetry operation
is shifted by $\bm{d}=\bm{d}_{\perp}+\bm{d}_{\parallel}$,
the nonsymmorphic mirror symmetry $\left\{M| \bm{t} \right\}$ also translates
to $\left\{M| \bm{t}-2\bm{d}_{\perp}\right\}$.
Thus by choosing $2\bm{d}_{\perp}=\bm{t}_{\perp}$,
the perpendicular component of the partial translation can be erased.
The resulting nonsymmorphic mirror symmetry
is conventionally considered as the definition of a glide mirror symmetry
$\widetilde{M}^{\parallel}\equiv\left\{M| \bm{t}_{\parallel} \right\}$.

However, $\bm{t}_{\perp}$ can also play a nontrivial role
in the presence of an additional point group symmetry $\{g|\bm{t'}\}$ 
centered at a different reference point with $\bm{t'}_{\perp}\neq\bm{t}_{\perp}$
modulo unit lattice translation. 
For instance, one can choose the inversion center as the reference point of the point group symmetry, thus inversion is given by $\{P|\bm{0}\}$
whereas the nonsymmorphic mirror is
$\left\{M| \bm{t} \right\}$.
Here the important point is that an additional shift of the reference point affects the form of the two operators
simultaneously. Namely, under the shift of the reference point by $\bm{d}=\bm{d}_{\perp}+\bm{d}_{\parallel}$,
the two symmetry operators transform as
$\left\{M| \bm{t} \right\}\longrightarrow \left\{M| \bm{t}-2\bm{d}_{\perp} \right\}$ and
$\left\{P| \bm{0} \right\}\longrightarrow \left\{P| -2\bm{d}_{\perp}-2\bm{d}_{\parallel} \right\}$,
which indicates that even if $\bm{t}_{\perp}$ is subtracted from the nonsymmorphic mirror symmetry by choosing
$2\bm{d}_{\perp}=\bm{t}_{\perp}$, it preserves its identity
in conjunction with the inversion symmetry $P$.
Therefore in systems with the inversion symmetry,
an off-centered mirror symmetry, defined as
\begin{align}
\widetilde{M}^{\perp}\equiv\left\{M| \bm{t}_{\perp} \right\},
\end{align}
deserves a separate consideration.

An off-centered rotation symmetry
can also be defined in a similar way.
A generic nonsymmorphic rotation symmetry element $\widetilde{C}_{n}=\left\{C_{n}| \bm{t} \right\}$ ($n=2,3,4,6$)
satisfies
\begin{align}
C_{n}\bm{t}_{\parallel}=\bm{t}_{\parallel},\quad C_{n}\bm{t}_{\perp}=\bm{t}'_{\perp},
\end{align}
where $C_{n}$ denotes the $n$-fold rotation symmetry
and $\bm{t}'_{\perp}$ is a partial translation rotated by $C_{n}$
satisfying $\bm{t}_{\perp}\cdot\bm{t}'_{\perp}=|\bm{t}_{\perp}|^{2}\cos\frac{2\pi}{n}$.
Since $C_{n}$ fulfills
$C_{n}^{n}=-1$ ($C_{n}^{n}=+1$) for particles with a half-integer (integer) spin,
a nonsymmorphic rotation symmetry $\left\{C_{n}| \bm{t} \right\}$ is under the following constraint,
$\left\{C_{n}| \bm{t} \right\}^{n}=\left\{C_{n}^{n}| n\bm{t}_{\parallel} \right\}\in \mathbb{T}$,
thus $\bm{t}_{\parallel}$ should have the form of $\bm{t}_{\parallel}=\frac{p}{n}\hat{a}_{\parallel}$ $p=0,1,...,n-1$
where $\hat{a}_{\parallel}$ is the unit translation along the rotation axis.
Again, $\bm{t}_{\perp}$ is not constrained in this case.

If the reference point for the point group symmetry operation
is shifted by $\bm{d}=\bm{d}_{\perp}+\bm{d}_{\parallel}$,
the nonsymmorphic rotation symmetry $\{C_{n}|\bm{t}\}$
also transforms to $\{C_{n}|\bm{t}+C_{n}\bm{d}_{\perp}-\bm{d}_{\perp}\}$.
Thus by choosing $\bm{d}_{\perp}$ to satisfy $\bm{t}_{\perp}=\bm{d}_{\perp}-C_{n}\bm{d}_{\perp}$,
$\bm{t}_{\perp}$ can be removed, leading
to a conventional screw rotation symmetry $\widetilde{C}_{n}^{\parallel}\equiv\{C_{n}|\bm{t}_{\parallel}\}$.
However, in the presence of an additional point group symmetry centered
at a different reference point, such as $\{P|\bm{0}\}$,
an off-centered nonsymmorphic rotation symmetry
\begin{align}
\widetilde{C}_{n}^{\perp}\equiv\{C_{n}|\bm{t}_{\perp}\}
\end{align}
can be defined,
and the partial translation $\bm{t}_{\perp}$ can cause intriguing physical consequences as shown
in the following.

{\bf Point nodes protected by off-centered rotation symmetries.}

In electronic systems having both time-reversal and
inversion symmetries,
eigenstates are doubly degenerate at any momentum.
Due to level repulsion between degenerate
bands, accidental band degeneracy is lifted
unless additional crystalline symmetry is supplemented~\cite{Yang_classification1}.
Here we show that the presence of an off-centered symmetry
creates symmetry-protected band degeneracy at the BZ boundary.
For simplicity, let us first introduce an off-centered two-fold rotation $\widetilde{C}^{\perp}_{2z}=\{C_{2z}|\frac{1}{2}\hat{x}+\frac{1}{2}\hat{y}\}$
to an orthorhombic crystal with $T$ and $P$ symmetries.
Here $\hat{x}$, $\hat{y}$, $\hat{z}$ denote the unit lattice vectors in the $x, y, z$ directions, respectively.
To understand the origin of band degeneracy,
let us examine how a spatial coordinate $\bm{r}=(x,y,z)$ transforms under $\widetilde{C}^{\perp}_{2z}$, 
\begin{align}\label{eqn:C2z_perp}
\widetilde{C}^{\perp}_{2z}~:&~ (x,y,z)\longrightarrow (-x+\frac{1}{2},-y+\frac{1}{2},z),
\nonumber\\
\left[\widetilde{C}^{\perp}_{2z}\right]^{2}~:&~ (x,y,z)\longrightarrow (x,y,z).
\end{align}
One can see that since $\left[\widetilde{C}^{\perp}_{2z}\right]^{2}$ does not accompany a partial translation,
it is actually equivalent to a symmorphic operation $C_{2z}^{2}$,
which leads to $\left[\widetilde{C}^{\perp}_{2z}\right]^{2}=-1$ independent of the spatial coordinate.
Thus at the momentum $\bm{k}$ invariant under $\widetilde{C}^{\perp}_{2z}$, each band $|\Psi(\bm{k})\rangle$
can be labelled by the momentum independent $\widetilde{C}^{\perp}_{2z}$ eigenvalue $\pm i$,
$\widetilde{C}^{\perp}_{2z}|\Psi_{\pm}(\bm{k})\rangle=\pm i|\Psi_{\pm}(\bm{k})\rangle$.
Since the system is invariant under $\widetilde{C}^{\perp}_{2z}$
along the four lines $\bm{k}_{1}=(0,0,k_{z})$,
$\bm{k}_{2}=(\pi,0,k_{z})$, $\bm{k}_{3}=(0,\pi,k_{z})$, $\bm{k}_{4}=(\pi,\pi,k_{z})$ with $k_{z}\in [-\pi,\pi]$,
a state $|\Psi(\bm{k})\rangle$ on any of these lines carries a constant $\widetilde{C}^{\perp}_{2z}$ eigenvalue.

This can be contrasted to the case of 
a two-fold screw rotation $\widetilde{C}^{\parallel}_{2z}=\{C_{2z}|\frac{1}{2}\hat{z}\}$ satisfying
\begin{align}\label{eqn:C2z_parallel}
\left[\widetilde{C}^{\parallel}_{2z}\right]^{2}~:&~ (x,y,z)\longrightarrow (x,y,z+1).
\end{align}
Along the line invariant under $\widetilde{C}^{\parallel}_{2z}$, the relevant eigenstates
satisfy
$\widetilde{C}^{\parallel}_{2z}|\Psi_{\pm}(\bm{k})\rangle=\pm ie^{i\frac{1}{2}k_{z}}|\Psi_{\pm}(\bm{k})\rangle$.
Due to the momentum dependence of the eigenvalues, the two different $\widetilde{C}^{\parallel}_{2z}$ eigensectors should be interchanged
when the momentum $k_{z}$ is shifted by $2\pi$.

Now we consider the combined effect of $P$ and $\widetilde{C}^{\perp}_{2z}$.
From the combined transformations
\begin{align}
P\widetilde{C}^{\perp}_{2z}~:&~ (x,y,z)\longrightarrow (x-\frac{1}{2},y-\frac{1}{2},-z),
\nonumber\\
\widetilde{C}^{\perp}_{2z}P~:&~ (x,y,z)\longrightarrow (x+\frac{1}{2},y+\frac{1}{2},-z),
\end{align}
we obtain
\begin{align}
\widetilde{C}^{\perp}_{2z}P|\Psi(\bm{k})\rangle=e^{ik_{x}+ik_{y}}P\widetilde{C}^{\perp}_{2z}|\Psi(\bm{k})\rangle.
\end{align}
Thus along the two $\widetilde{C}^{\perp}_{2z}$ invariant lines
$\bm{k}_{2}=(\pi,0,k_{z})$, $\bm{k}_{3}=(0,\pi,k_{z})$ with $k_{z}\in [-\pi,\pi]$,
$P$ and $\widetilde{C}^{\perp}_{2z}$ anticommute, i.e., $\{\widetilde{C}^{\perp}_{2z},P\}=0$. Moreover,
since the time-reversal symmetry $T$ commutes with both $P$ and $\widetilde{C}^{\perp}_{2z}$,
we obtain
$\{\widetilde{C}^{\perp}_{2z},PT\}=0$,
which gives rise to
\begin{align}
\widetilde{C}^{\perp}_{2z}\left[PT|\Psi_{\pm}(\bm{k})\rangle\right]&=-PT\left[\widetilde{C}^{\perp}_{2z}|\Psi_{\pm}(\bm{k})\rangle\right]
=-PT\left[\pm i|\Psi_{\pm}(\bm{k})\rangle\right]
=\pm i\left[PT|\Psi_{\pm}(\bm{k})\rangle\right].
\end{align}
Thus $|\Psi_{\pm}(\bm{k})\rangle$ and $PT|\Psi_{\pm}(\bm{k})\rangle$,
which are locally degenerate at the momentum $\bm{k}$,
have the same $\widetilde{C}^{\perp}_{2z}$ eigenvalues of $\pm i$.
Therefore when two degenerate bands having different
$\widetilde{C}^{\perp}_{2z}$ eigenvalues cross,
the resulting band crossing point is protected and forms a 3D Dirac point
with four-fold degeneracy.

For comparison, let us consider a similar problem in systems with a two-fold screw rotation
$\widetilde{C}^{\parallel}_{2z}=\{C_{2z}|\frac{1}{2}\hat{z}\}$.
It is straightforward to show that
$\widetilde{C}^{\parallel}_{2z}P|\Psi(\bm{k})\rangle=e^{ik_{z}}P\widetilde{C}^{\parallel}_{2z}|\Psi(\bm{k})\rangle$.
Then along the line invariant under $\widetilde{C}^{\parallel}_{2z}$,
where $\widetilde{C}_{2z}^{\parallel}|\Psi_\pm(\bm{k})\rangle=\pm ie^{-\frac{i}{2}k_z}|\Psi_\pm(\bm{k})\rangle$, we obtain
\begin{align}
\widetilde{C}^{\parallel}_{2z}\!\left[PT|\Psi_{\pm}(\bm{k})\rangle\right]
&=e^{-ik_z}PT\!\left[\widetilde{C}^{\parallel}_{2z}
|\Psi_{\pm}(\bm{k})\rangle\right]
=e^{-ik_z}PT\!\left[\pm ie^{-\frac{i}{2}k_z}|\Psi_{\pm}(\bm{k})\rangle\right]
=\mp ie^{-\frac{i}{2}k_z}PT|\Psi_{\pm}(\bm{k})\rangle,
\end{align}
which show that the degenerate states $|\Psi_{\pm}(\bm{k})\rangle$ and $PT|\Psi_{\pm}(\bm{k})\rangle$ belong
to different eigensectors of $\widetilde{C}^{\parallel}_{2z}$ symmetry.
Therefore, when two bands, each of which is doubly degenerate, touch,
there always is some finite hybridization between degenerate bands.
Thus $\widetilde{C}^{\parallel}_{2z}$ symmetry cannot protect a stable Dirac point
at a generic momentum.
One exception is when the band crossing happens at the time-reversal invariant momentum (TRIM) with $k_{z}=\pi$.
In this case, two bands having the same $\widetilde{C}^{\parallel}_{2z}$ eigenvalues form
a Kramers pair, and
two Kramers pairs having different $\widetilde{C}^{\parallel}_{2z}$ eigenvalues are connected by $P$, leading to four-fold degeneracy~\cite{YoungKane_2D}. 
However, such a degeneracy point does not form a 3D Dirac point.
Instead, it becomes a part of a line node in the $k_z=\pi$ plane protected by $\widetilde{M}^{\perp}_z=\widetilde{C}^{\parallel}_{2z}P$, as discussed in the following.

In fact, the anticommutation relation between $P$ and $\widetilde{C}^{\perp}_{2z}$
puts a strong constraint on the band structure along
the $\widetilde{C}^{\perp}_{2z}$ invariant axis.
Considering
\begin{align}
\widetilde{C}^{\perp}_{2z}\left[P|\Psi_{\pm}(\bm{k})\rangle\right]&=-P\left[\widetilde{C}^{\perp}_{2z}|\Psi_{\pm}(\bm{k})\rangle\right]
=-P\left[\pm i|\Psi_{\pm}(\bm{k})\rangle\right]
=\mp i\left[P|\Psi_{\pm}(\bm{k})\rangle\right],
\end{align}
one can find that two energetically degenerate states $|\Psi_{\pm}(\bm{k})\rangle$ and $P|\Psi_{\pm}(\bm{k})\rangle$,
which are located at $\bm{k}$ and $-\bm{k}$, respectively,
have the opposite $\widetilde{C}^{\perp}_{2z}$ eigenvalues.
Let us recall that at each momentum $\bm{k}$, a Kramers pair should have the same $\widetilde{C}^{\perp}_{2z}$ eigenvalue.
This means that on the $\widetilde{C}^{\perp}_{2z}$ invariant axis where $\{P,\widetilde{C}^{\perp}_{2z}\}=0$ is satisfied,
there should be a pair of degenerate bands with different $\widetilde{C}^{\perp}_{2z}$ eigenvalues,
which we call {\it a doublet pair}.
Since a doublet pair should form a band structure which is symmetric with respect to a TRIM,
they should be degenerate at the two TRIMs on the $\widetilde{C}^{\perp}_{2z}$ invariant axis as shown in Fig.~\ref{fig:typeII_rotation} (a) and (b).
Here each of the degenerate points with four-fold degeneracy represent a
3D Dirac point located at a TRIM.

Due to the presence of a quantized $\widetilde{C}^{\perp}_{2z}$ eigenvalue,
the band crossing points between two different doublet pairs can also generate
3D Dirac points. Namely,
as long as the two crossing bands have different
$\widetilde{C}^{\perp}_{2z}$ eigenvalues,
the crossing points are symmetry protected.
In general, such a crossing between doublet pairs
generates $4n$ ($n$ is an integer) band crossing points, and the location of
each Dirac point is away from TRIMs as shown in Fig.~\ref{fig:typeII_rotation}(c).

{\bf Line nodes protected by off-centered mirror symmetries.}

An off-centered mirror symmetry can create a stable line node with four-fold degeneracy
in systems with $P$ and $T$ symmetries.
For convenience, let us consider $\widetilde{M}^{\perp}_{x}=\{M_{x}|\frac{1}{2}\hat{x}\}$,
which transforms a spatial coordinate $\bm{r}$ in the following way,
\begin{align}\label{eqn:Mx_perp}
\widetilde{M}^{\perp}_{x}~&:~(x,y,z)\rightarrow (-x+\frac{1}{2},y,z),
\nonumber\\
\left[\widetilde{M}^{\perp}_{x}\right]^{2}~&:~(x,y,z)\rightarrow (x,y,z).
\end{align}
From $M_{x}^{2}=-1$, we obtain $\left[\widetilde{M}^{\perp}_{x}\right]^{2}=-1$ independent of a spatial coordinate.
Thus at the momentum $\bm{k}$ invariant under $\widetilde{M}^{\perp}_{x}$,
i.e., at any momentum in the 2D plane with $k_{x}=0$ or $k_{x}=\pi$,
each band $|\Psi(\bm{k})\rangle$ can be labelled by the momentum independent
$\widetilde{M}^{\perp}_{x}$ eigenvalue $\pm i$, i.e.,
$\widetilde{M}^{\perp}_{x}|\Psi_{\pm}(\bm{k})\rangle
=\pm i|\Psi_{\pm}(\bm{k})\rangle$.

Let us compare this to the case of a glide mirror
$\widetilde{M}^{\parallel}_{x}=\{M_{x}|\frac{1}{2}\hat{y}+\frac{1}{2}\hat{z}\}$ satisfying
\begin{align}\label{eqn:Mx_parallel}
\left[\widetilde{M}^{\parallel}_{x}\right]^{2}~:&~ (x,y,z)\longrightarrow (x,y+1,z+1).
\end{align}
In a plane invariant under $\widetilde{M}^{\parallel}_{x}$, the eigenstates
satisfy
$\widetilde{M}^{\parallel}_{x}|\Psi_{\pm}(\bm{k})\rangle=
\pm ie^{\frac{i}{2}(k_{y}+k_{z})}|\Psi_{\pm}(\bm{k})\rangle$.
Due to the momentum dependence of the eigenvalues, the two different $\widetilde{M}^{\parallel}_{x}$ eigensectors should be interchanged
when either $k_{y}$ or $k_{z}$ is shifted by $2\pi$.

Now we consider the combined effect of $P$ and $\widetilde{M}^{\perp}_{x}$.
From
\begin{align}
P\widetilde{M}^{\perp}_{x}~:&~ (x,y,z)\longrightarrow (x-\frac{1}{2},-y,-z),
\nonumber\\
\widetilde{M}^{\perp}_{x}P~:&~ (x,y,z)\longrightarrow (x+\frac{1}{2},-y,-z),
\end{align}
we obtain
\begin{align}
\widetilde{M}^{\perp}_{x}P|\Psi(\bm{k})\rangle=e^{ik_{x}}P\widetilde{M}^{\perp}_{x}|\Psi(\bm{k})\rangle.
\end{align}
Thus in the $k_{x}=\pi$ plane,
$P$ and $\widetilde{M}^{\perp}_{x}$ anticommute, i.e., $\{\widetilde{M}^{\perp}_{x},P\}=0$. Moreover,
since the time-reversal symmetry $T$ commutes with both $P$ and $\widetilde{M}^{\perp}_{x}$,
we obtain
$\{\widetilde{M}^{\perp}_{x},PT\}=0$,
which gives rise to
\begin{align}
\widetilde{M}^{\perp}_{x}\left[PT|\Psi_{\pm}(\bm{k})\rangle\right]&=-PT\left[\widetilde{M}^{\perp}_{x}|\Psi_{\pm}(\bm{k})\rangle\right]
=-PT\left[\pm i|\Psi_{\pm}(\bm{k})\rangle\right]
=\pm i\left[PT|\Psi_{\pm}(\bm{k})\rangle\right].
\end{align}
Thus $|\Psi_{\pm}(\bm{k})\rangle$ and $PT|\Psi_{\pm}(\bm{k})\rangle$, which are degenerate at the momentum $\bm{k}$,
have the same $\widetilde{M}^{\perp}_{x}$ eigenvalues of $\pm i$.
Therefore when two degenerate bands having different $\widetilde{M}^{\perp}_{x}$ eigenvalues cross,
the resulting band crossing point is protected and forms a line node
with four-fold degeneracy on the invariant plane $k_x=\pi$.

For comparison, let us consider a similar problem in systems with a glide mirror
$\widetilde{M}^{\parallel}_{x}=\{M_{x}|\frac{1}{2}\hat{y}+\frac{1}{2}\hat{z}\}$.
It is straightforward to show that
\begin{align}
\widetilde{M}^{\parallel}_{x}P|\Psi(\bm{k})\rangle=e^{i(k_{y}+k_{z})}P\widetilde{M}^{\parallel}_{x}|\Psi(\bm{k})\rangle.
\end{align}
Then in a 2D plane invariant under $\widetilde{M}^{\parallel}_{x}$, we obtain
\begin{align}
\widetilde{M}^{\parallel}_{x}\left[PT|\Psi_{\pm}(\bm{k})\rangle\right]
&=e^{-i(k_{y}+k_{z})}PT\left[\widetilde{M}^{\parallel}_{x}|\Psi_{\pm}(\bm{k})\rangle\right]
=\mp ie^{-\frac{i}{2}(k_{y}+k_{z})}\left[PT|\Psi_{\pm}(\bm{k})\rangle\right],
\end{align}
which shows that $|\Psi_{\pm}(\bm{k})\rangle$ and $PT|\Psi_{\pm}(\bm{k})\rangle$,
which are degenerate at the momentum $\bm{k}$, belong
to different eigensectors of $\widetilde{M}^{\parallel}_{x}$ symmetry.
This means that when two bands, each doubly degenerate due to
the $PT$ symmetry, overlap,
there always is some finite hybridization between them 
at a generic momentum, thus a stable line node cannot be protected by $\widetilde{M}^{\parallel}_{x}$ symmetry in a mirror invariant plane. Instead, stable Dirac point nodes are protected by an off-centered symmetry $\widetilde{C}^{\perp}_{2x}=\widetilde{M}^{\parallel}_{x} P$ on its invariant lines $(k_x,\pi,0)$ and $(k_x,0,\pi)$.


In fact, the anticommutation relation between $P$ and $\widetilde{M}^{\perp}_{x}$
puts a strong constraint on the band structure in the $\widetilde{M}^{\perp}_{x}$ invariant plane.
Considering
\begin{align}
\widetilde{M}^{\perp}_{x}\left[P|\Psi_{\pm}(\bm{k})\rangle\right]&=-P\left[\widetilde{M}^{\perp}_{x}|\Psi_{\pm}(\bm{k})\rangle\right]
=-P\left[\pm i|\Psi_{\pm}(\bm{k})\rangle\right]
=\mp i\left[P|\Psi_{\pm}(\bm{k})\rangle\right],
\end{align}
we find that two energetically degenerate states $|\Psi_{\pm}(\bm{k})\rangle$ and $P|\Psi_{\pm}(\bm{k})\rangle$,
which are located at $\bm{k}$ and $-\bm{k}$, respectively,
have the opposite $\widetilde{M}^{\perp}_{x}$ eigenvalues.
It is worth to remind that a Kramers pair at each momentum $\bm{k}$, which are degenerate due to $PT$ symmetry,
have the same $\widetilde{M}^{\perp}_{x}$ eigenvalue.
This means that in the $k_{x}=\pi$ plane where $\{P,\widetilde{M}^{\perp}_{x}\}=0$ is satisfied,
two bands (each with Kramers degeneracy) having different $\widetilde{M}^{\perp}_{x}$ eigenvalues
should form {\it a doublet pair} again as in the case of
the off-centered rotation symmetry.
Since the whole band structure in the $k_{x}=\pi$ plane is symmetric with respect to a TRIM,
each doublet pair should be degenerate along a line which passes two TRIMs as shown in Fig.~\ref{fig:typeII_mirror}(a) and (b).
Here a set of the degenerate points form
a line node with four-fold degeneracy.

Due to the presence of quantized $\widetilde{M}^{\perp}_{x}$ eigenvalues,
a band crossing between two different doublet pairs can also generate
nodal lines. Namely,
as long as the two bands have different $\widetilde{M}^{\perp}_{x}$ eigenvalues,
their crossing points are symmetry protected.
In general, such crossing between two different doublet pairs
generate $4n$ ($n$ is an integer) nodal lines, and the location of
each nodal line is away from TRIM as shown in
Fig.~\ref{fig:typeII_mirror}(c).

{\bf Model.}

\begin{figure*}[t]
\centering
\includegraphics[width=16 cm]{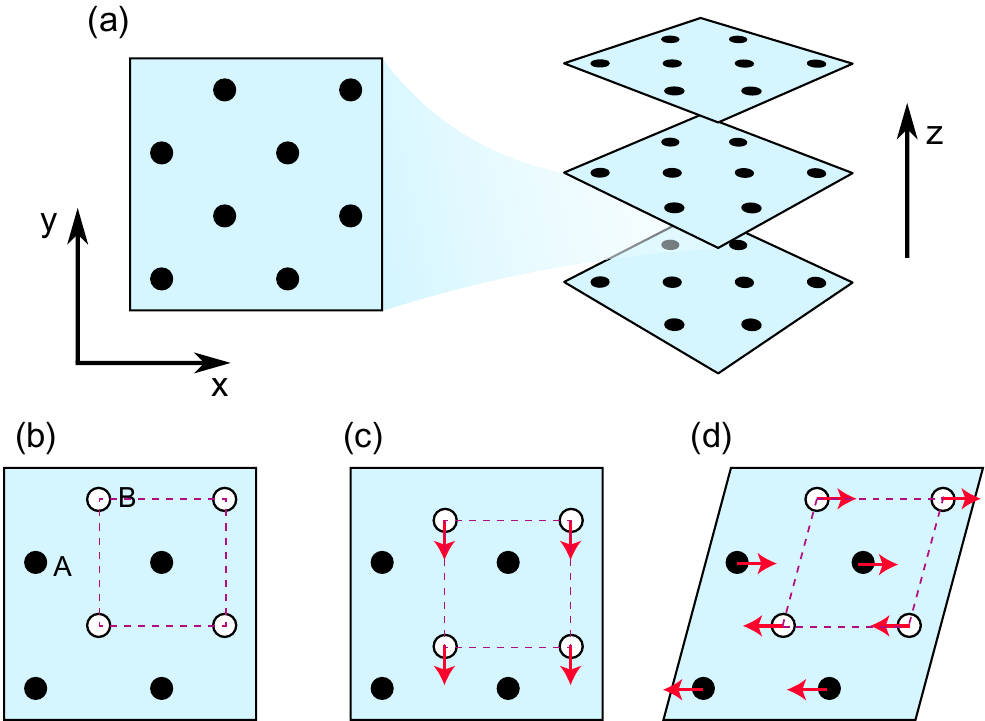}
\caption{
{\bf Construction of 3D lattice models by stacking 2D layers.}
({\bf a})
A schematic figure describing a 3D lattice model obtained by vertical stacking of 2D square lattices.
({\bf b})
Structure of a 2D layer
where the B site in a unit cell is shifted along the $z$-direction, thus
the whole system has nonsymmorphic symmetries.
({\bf c})
An additional shifting of B sites along the $y$-direction, which breaks
the symmetry $\widetilde{C}^{\perp}_{2z}$ (or equivalently, $\widetilde{M}^{\parallel}_{z}$).
({\bf d})
An additional distortion of a unit cell, which breaks
the symmetry $\widetilde{M}^{\perp}_{x}$ and $\widetilde{M}^{\perp}_{y}$
(or equivalently, $\widetilde{C}^{\parallel}_{2x}$ and $\widetilde{C}^{\parallel}_{2y}$).
} \label{fig:lattice_structure}
\end{figure*}

To demonstrate the general idea discussed up to now,
we construct a 3D tight-binding Hamiltonian on a tetragonal lattice,
which is composed of 2D square lattices stacked along the $z$-direction as described in Fig.~\ref{fig:lattice_structure}.
For a 2D layer, we adopt the lattice model proposed in Ref.~\onlinecite{YoungKane_2D}
in which a unit cell contains two sublattice sites, labeled $A$ and $B$,
where the $B$ sublattice is displaced by
$\bm{r}_{AB}=(\frac{1}{2},\frac{1}{2},\delta_{z})$ ($0<\delta_{z}<1$)
from the $A$ sublattice.
Here we assume that both the in-plane and out-of-plane lattice constants to be unity.
The vertical shift $\delta_{z}$ makes the symmetry of the lattice to be non-symmorphic.
Explicitly,
the Hamiltonian in the real space is given by
\begin{align}\label{eqn:model_Rspace}
\hat{H}^{(0)}=\sum_{\langle i,j\rangle}t(\bm{r}_{ij})\hat{c}^{\dag}_{\bm{r}_{i}}\hat{c}_{\bm{r}_{j}}
+\sum_{\langle\langle i,j\rangle\rangle}t'(\bm{r}_{ij})\hat{c}^{\dag}_{\bm{r}_{i}}\hat{c}_{\bm{r}_{j}}
+\sum_{\langle i,j,k\rangle}i\lambda(\bm{r}_{ij},\bm{r}_{jk})\hat{c}^{\dag}_{\bm{r}_{i}}
\left[\left(\bm{r}_{ij}\times\bm{r}_{jk}\right)\cdot\bm{\sigma}\right]\hat{c}_{\bm{r}_{k}}
\end{align}
where $t(\bm{r}_{ij})$ ($t'(\bm{r}_{ij})$) is the hopping amplitude
between same (different) sublattice sites, and
$\lambda(\bm{r}_{ij},\bm{r}_{jk})$ denotes the spin-orbit induced hopping amplitude between the same sublattice sites $i$ and $k$
through the site $j$ belonging to the other sublattice.
Here $\bm{r}_{ij}=\bm{r}_{i}-\bm{r}_{j}$ and
the Pauli matrix $\bm{\sigma}$ indicates the spin degrees of freedom.
More detailed information about the lattice model is given in Methods.

Let us note that $\hat{H}^{(0)}$ possesses not only 
the time-reversal symmetry $T$ and
the inversion symmetry $P$ but also the off-centered symmetries
$\widetilde{C}^{\perp}_{2z}=\{C_{2z}|\frac{1}{2}\hat{x}+\frac{1}{2}\hat{y}\}$,
$\widetilde{M}^{\perp}_{x}=\{M_{x}|\frac{1}{2}\hat{x}\}$, and
$\widetilde{M}^{\perp}_{y}=\{M_{y}|\frac{1}{2}\hat{y}\}$, thus the system
corresponds to the space group No.59.
The corresponding screw/glide symmetries can be defined as
$\widetilde{M}^{\parallel}_{z}=\widetilde{C}^{\perp}_{2z}P=\{M_{z}|\frac{1}{2}\hat{x}+\frac{1}{2}\hat{y}\}$,
$\widetilde{C}^{\parallel}_{2x}=\widetilde{M}^{\perp}_{x}P=\{C_{2x}|\frac{1}{2}\hat{x}\}$, and
$\widetilde{C}^{\parallel}_{2y}=\widetilde{M}^{\perp}_{y}P=\{C_{2y}|\frac{1}{2}\hat{y}\}$.
By shifting the location of the $B$ site relative to the $A$ site in a unit cell,
the symmetry of the Hamiltonian can be systematically lowered, thus one can
examine the role of a particular symmetry to protect a relevant semimetal phase using a single lattice model.

First, we shift the position of the B site in a unit cell in the $y$ direction,
which makes $\bm{r}_{AB}=(\frac{1}{2},\delta_{y}\neq\frac{1}{2},\delta_{z})$ as shown in Fig.~\ref{fig:lattice_structure}(c).
This distortion breaks $\widetilde{C}^{\perp}_{2z}$ and $\widetilde{M}^{\perp}_{y}$ symmetries,
whereas $\widetilde{M}^{\perp}_{x}$ is preserved as well as the $P$ and $T$ symmetries,
thus the system corresponds to the space group No.11.
The resulting band structure is shown in Fig.~\ref{fig:linenode_doubling}(a).
One can clearly see that there are two line nodes in the $k_{x}=\pi$ plane,
and each line node connects two TRIMs, which is consistent with the prediction of the general theory.
To observe the band crossing between two doublet pairs described in Fig.~\ref{fig:typeII_mirror}(c),
we construct an eight-band model by adding two copies of the $4\times 4$ Hamiltonian in Eq.~(\ref{eqn:model_Rspace}).
As the hybridization between the two $4\times 4$ blocks is turned on,
each line node passing two TRIMs splits into two different nodal lines,
thus one can observe four nodal lines, and none of them passes a TRIM
as shown in Fig.~\ref{fig:linenode_doubling}(b)
and Fig.~\ref{fig:typeII_mirror}(c).

The second distortion is achieved by deforming the lattice along the [110] direction,
which breaks $\widetilde{M}^{\perp}_{x}$ and
$\widetilde{M}^{\perp}_{y}$ symmetries
whereas $\widetilde{C}^{\perp}_{2z}$ is preserved as well as the $P$ and $T$ symmetries,
thus the system corresponds to the space group No.13.
(See Fig.~\ref{fig:lattice_structure} (d).)
As shown in Fig.~\ref{fig:pointnode_doubling} (a), one can observe four Dirac points
protected by $\widetilde{C}^{\perp}_{2z}$ symmetry
located at TRIMs $\bm{k}=(\pi,0,0)$, $(\pi,0,\pi)$, $(0,\pi,0)$,
and $(0,\pi,\pi)$.
When the number of bands is doubled by combining two different $4\times 4$ Hamiltonians,
one can observe four Dirac points on the line $\bm{k}=(\pi,0,k_{z})$ and also $\bm{k}=(0,\pi,k_{z})$
with $k_{z}\in(-\pi,\pi)$, respectively, as shown in Fig.~\ref{fig:pointnode_doubling} (b).
Here none of Dirac points is located at a TRIM in agreement with the prediction of the general theory,
and the relevant band structure is also consistent with Fig.~\ref{fig:typeII_rotation} (c).

{\bf Time-reversal symmetry breaking and Fermi surface topology.}

Since the stability of a Dirac point/line node
requires the simultaneous presence of an off-centered
symmetry ($\widetilde{M}^{\perp}_{x}$ or $\widetilde{C}^{\perp}_{2z}$)
together with the $T$ and $P$,
it is interesting to examine the influence of symmetry breaking on the band structure.
In particular, we find that the breaking of time reversal symmetry, 
due to Zeeman effect from external magnetic field ($\bm{H}$)
or exchange coupling from doped magnetic ions,
can create intriguing evolution in both the bulk and surface band structures,
as summarized in Table I.

\begin{figure*}[t]
\centering
\includegraphics[width=16 cm]{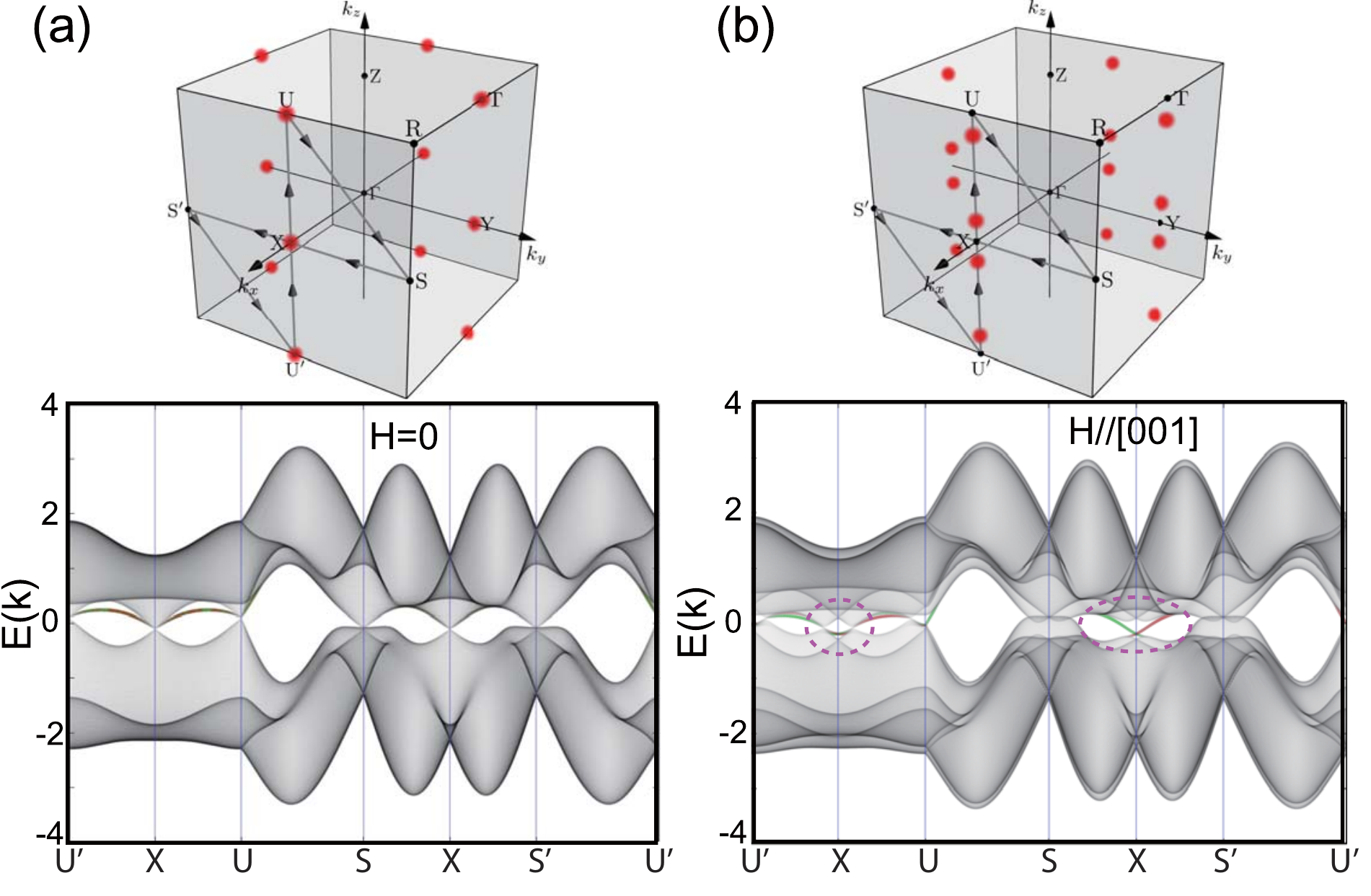}
\caption{
{\bf Magnetic field induced transition 
of Dirac points into Weyl points.}
The bulk band structure of the 4-band lattice model
with $P$, $T$, and $\widetilde{C}^{\perp}_{2z}$ symmetries is shown
in the lower panels after projection onto the surface BZ  of
a slab structure with a finite length $L_x$
along the $x$-direction.
The states localized on the $x=0$ ($x=L_x$) surface
are indicated by red (green) lines, respectively.
({\bf a})
In the absence of magnetic field. There are four Dirac points at TRIMs
(X, U, Y, T)
at $\bm{k}=(\pi,0,0)$, $(\pi,0,\pi)$,
$(0,\pi,0)$, $(0,\pi,\pi)$, respectively. 
There are non-topological surface states on both surfaces
which can be merged into bulk states through smooth deformation.
({\bf b})
In the presence of magnetic field along the $z$-direction ($h_z=0.2$).
A Dirac point with four-fold degeneracy splits into two Weyl points,
each with two-fold degeneracy.
Splitting of a Dirac point along the $k_{z}$-direction 
accompanies emergent surface states (Fermi arcs)
connecting two split Weyl points,
which are marked with red dotted circles.
In the figures, $U'$ and $S'$ indicate the momenta equivalent to $U$ and $S$, respectively. 
} \label{fig:pointnode_Bfield}
\end{figure*}

In the case of the semimetal with Dirac point nodes protected by $\widetilde{C}^{\perp}_{2z}$ symmetry,
a Dirac point with four-fold degeneracy always splits into two Weyl points with two-fold degeneracy,
which accompanies a Fermi arc connecting the two Weyl points as shown in Fig.~\ref{fig:pointnode_Bfield}.
More specifically, when $\bm{H}\parallel\hat{\bm{z}}$, the system preserves $\widetilde{C}^{\perp}_{2z}$ symmetry, and
the two Weyl points split from a Dirac point is shifted along the $k_{z}$ direction.
On the other hand, when $\bm{H}\perp\hat{\bm{z}}$, the split Weyl points move in the plane normal to the $k_{z}$ direction.

\begin{figure*}[t]
\centering
\includegraphics[width=14 cm]{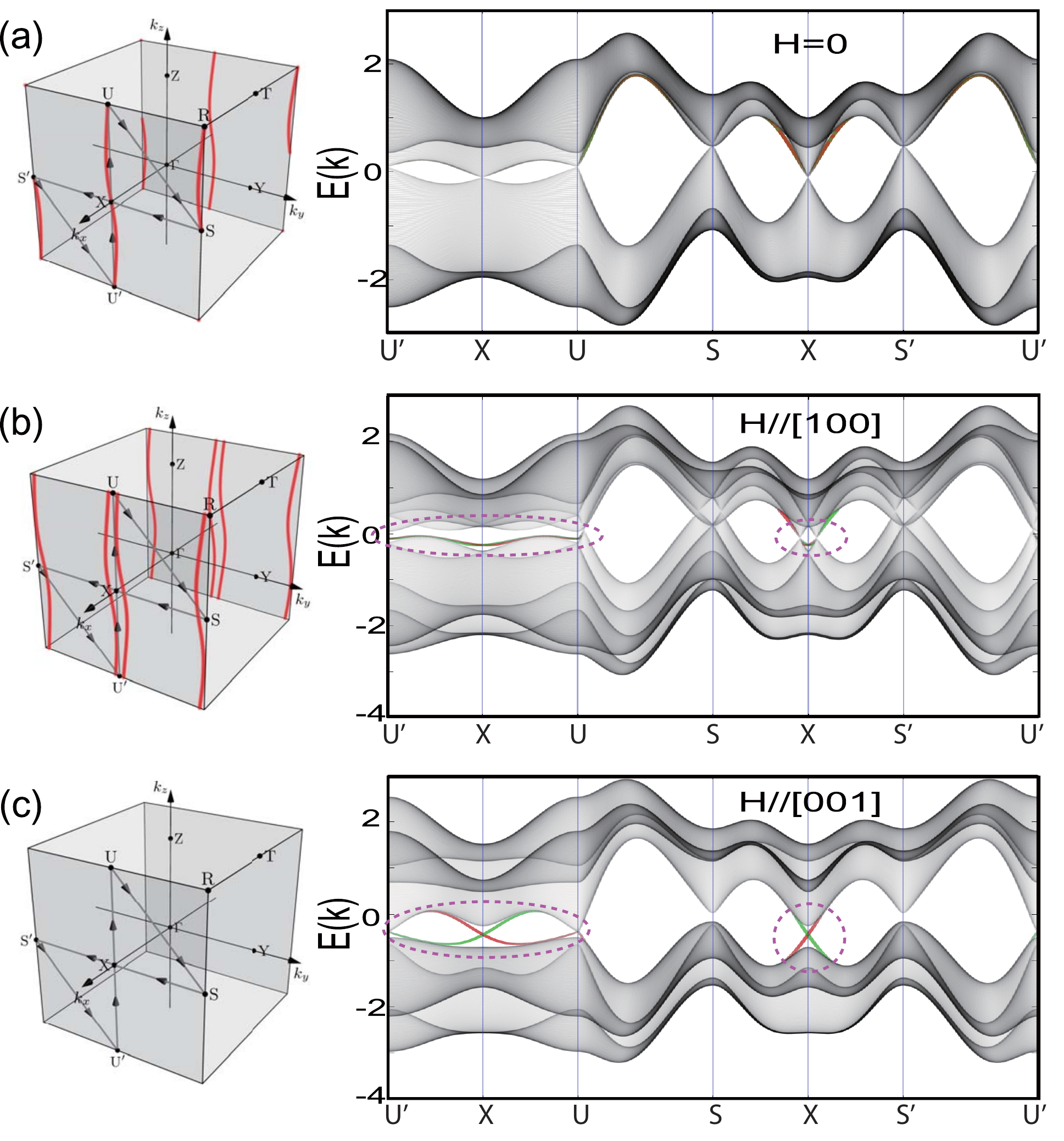}
\caption{
{\bf Magnetic field induced transition 
of Dirac line nodes into Weyl line nodes and three-dimensional
quantum Hall insulators.}
The bulk band structure of the 4-band lattice model
with $P$, $T$, and $\widetilde{M}^{\perp}_{x}$ symmetries is shown
in the right panels after projection onto the surface Brillouin zone of
a slab structure with a finite length $L_x$
along the $x$-direction.
The states localized on the $x=0$ ($x=L_x$) surface
are indicated by red (green) lines, respectively.
({\bf a})
In the absence of magnetic field. There are two Dirac line nodes on
the $k_{x}=\pi$ plane. There are some non-topological surface states
which can be merged with bulk states through smooth deformation.
({\bf b})
In the presence of magnetic field along the $x$-direction ($h_x=0.3$).
A Dirac line node with four-fold degeneracy splits into two Weyl line nodes,
each with two-fold degeneracy.
Splitting of a Dirac line node accompanies emergent surface states
connecting two split Weyl line nodes,
which are marked with red dotted circles.
({\bf c})
In the presence of magnetic field along the $z$-direction ($h_z=0.7$).
A Dirac line node is fully gapped,
and two-dimensional chiral surface states emerge,
which are marked with red dotted circles.
In the figures, $U'$ and $S'$ indicate the momenta equivalent to $U$ and $S$, respectively. 
} \label{fig:linenode_Bfield}
\end{figure*}

In the case of the semimetal with Dirac line nodes protected by $\widetilde{M}^{\perp}_{x}$ symmetry,
the application of the external magnetic field causes more dramatic physical consequences.
Firstly, when $\bm{H}\parallel\hat{\bm{x}}$, thus the systems preserves $\widetilde{M}^{\perp}_{x}$ symmetry,
a Dirac line node with four-fold degeneracy splits into
two Weyl line nodes with two-fold degeneracy as shown in Fig.~\ref{fig:linenode_Bfield}.
Here both the Dirac line node and the Weyl line nodes are located in the $k_{x}=\pi$ plane.
Interestingly, the splitting of a Dirac line node is accompanied by emergent 2D surface states
connecting the split Weyl line nodes, which originate from the $\pi$ Berry phase around each Weyl line node~\cite{inversion1,inversion2,Vanderbilt, Schnyder}.
The stability of the Weyl line node can be understood in the following way.
Since $\widetilde{M}^{\perp}_{x}$ symmetry is preserved in the whole $k_{x}=\pi$ plane even in the presence
of magnetic field ($\bm{H}\parallel\hat{\bm{x}}$),
each eigenstate still carries a quantized $\widetilde{M}^{\perp}_{x}$ eigenvalues of $\pm i$.
Moreover, due to the inversion symmetry $P$ satisfying $\{P,\widetilde{M}^{\perp}_{x}\}=0$,
if a state $|\Psi_{\pm}(\pi,k_{y},k_{z})\rangle$ at the momentum $\bm{k}=(\pi,k_{y},k_{z})$ has
the $\widetilde{M}^{\perp}_{x}$ eigenvalues of $\pm i$,
the state $|\Psi_{\pm}(\pi,-k_{y},-k_{z})\rangle\equiv P|\Psi_{\pm}(\pi,k_{y},k_{z})\rangle$ at the momentum $\bm{k}=(\pi,-k_{y},-k_{z})$ has
the $\widetilde{M}^{\perp}_{x}$ eigenvalues of $\mp i$.
Since the breaking of time-reversal symmetry splits each two-fold degenerate
band of zero field into two bands,
the band structure has a configuration similar to the one
shown in Fig.~\ref{fig:typeII_mirror}(c).
Let us note that a TRIM is invariant under the inversion as well.
Each state is non-degenerate at any $\bm{k}$ except at
TRIMs due to the broken time-reversal symmetry.
A stable Weyl line node with two-fold degeneracy is formed
as long as the degenerate states at the crossing point have
different $\widetilde{M}^{\perp}_{x}$ eigenvalues.

When $\bm{H}\perp\hat{\bm{x}}$ breaks
the $\widetilde{M}^{\perp}_{x}$ symmetry,
a Dirac line node with four-fold degeneracy is lifted,
and a band gap opens.
Figure~\ref{fig:linenode_Bfield} (c) shows the band structure
when $\bm{H}\parallel \hat{\bm{z}}$.
One can clearly see the opening of a band gap and the emergence
of chiral surface modes near the $U-X-U'$ line.
In this case, for any 2D $k_x$-$k_y$ plane with fixed $k_z$,
the Chern number of the bands below the gap is
equal to 1, and the system
can be viewed as a 3D quantum Hall insulator
having 2D chiral metallic states on the surface
(provided that the Fermi energy is in the band gap).
Therefore by changing the direction of the external magnetic field,
one can introduce a transition from a semimetal with line nodes
to a gapped phase, which can induce a dramatic change
in the magneto-transport properties.
All these results can also be confirmed by analyzing the low-energy
$\bm{k}\cdot\bm{p}$ Hamiltonian
as shown in detail in Methods.

\begin{figure*}[t]
\centering
\includegraphics[width=16 cm]{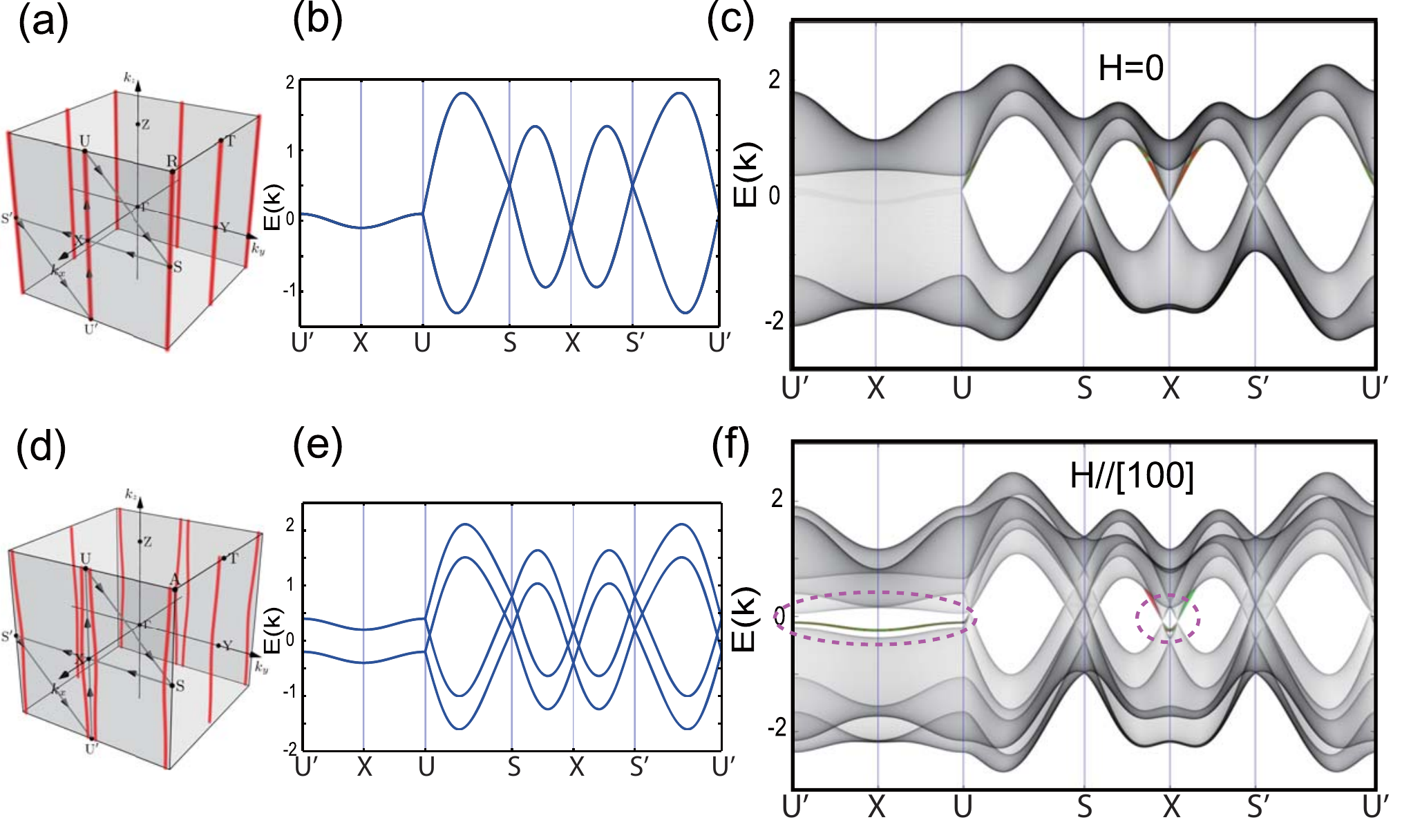}
\caption{
{\bf Magnetic field induced transition 
of Dirac line nodes into Weyl line nodes
in systems with both $\widetilde{M}^{\perp}_{x}$ and
$\widetilde{C}^{\perp}_{2z}$.}
({\bf a})
In the absence of magnetic field. There are three open straight Dirac line nodes
along $\bm{k}=(\pi,0,k_{z})$, $(0,\pi,k_{z})$ and $(\pi,\pi,k_{z})$ 
lines with $k_{z}\in[-\pi,\pi]$
due to the simultaneous presence of 
$\widetilde{M}^{\perp}_{x}$ and $\widetilde{M}^{\perp}_y$.
({\bf b}) The bulk band structure of the 4-band lattice model with
$P$, $T$, $\widetilde{C}^{\perp}_{2z}$,
and $\widetilde{M}^{\perp}_{x}$ symmetries.
({\bf c})
The band structure of the slab structure with a finite length
along the $x$-direction are shown.
The states localized on the $x=0$ ($x=L_x$) surface
are indicated by red (green) lines, respectively.
({\bf d}-{\bf f})
Similar plots 
in the presence of magnetic field along the $x$-direction ($h_x=0.3$).
A Dirac line node with four-fold degeneracy splits into two Weyl line nodes,
each with two-fold degeneracy.
Splitting of a Dirac line node accompanies emergent surface states connecting two split Weyl line nodes,
which are marked with dotted circles.
In the figures, $U'$ and $S'$ indicate the momenta equivalent to $U$ and $S$, respectively. 
} \label{fig:multiple_mirror}
\end{figure*}

Finally,
when two off-centered symmetries $\widetilde{C}^{\perp}_{2z}$ and $\widetilde{M}^{\perp}_{x}$ exist
at the same time, the system has an additional off-centered mirror symmetry $\widetilde{M}^{\perp}_{y}$
due to the relation $\widetilde{C}^{\perp}_{2z}=\widetilde{M}^{\perp}_{y}\widetilde{M}^{\perp}_{x}$.
The presence of multiple crystalline symmetries leads to
three Dirac line nodes at $\bm{k}=(\pi,0,k_{z})$,
$\bm{k}=(\pi,\pi,k_{z})$, $\bm{k}=(0,\pi,k_{z})$ with $k_{z}\in(-\pi,\pi)$,
each of which is an open line parallel to the $k_{z}$ direction as shown in Fig.~\ref{fig:multiple_mirror}.
When the magnetic field $\bm{H}\parallel[100]$ or
$\bm{H}\parallel[010]$ is applied to the system,
one can find a semimetal with Weyl line nodes since
at least one of the off-centered mirror symmetries is preserved in this case. 
One the other hand, when $\bm{H}\parallel[001]$, thus both of the off-centered mirror symmetries are broken, 
a gapped insulator appears as summarized in Table I.

\subsection*{Discussion}

In centrosymmetric crystals,
an off-centered two-fold rotation/mirror symmetry
is obtained as a product of a glide mirror/two-fold rotation
and inversion $P$.
Namely, $\widetilde{C}_2^{\perp}=\widetilde{M}^{\parallel}P$ and
$\widetilde{M}^{\perp}=\widetilde{C}_2^{\parallel}P$.
One can also ask about the role of other screw rotation symmetries
$\widetilde{C}_{n,p}^{\parallel}=\{C_{n}|\frac{p}{n}\hat{a}_{\parallel}\}$ ($n=3,4,6$ and $p=0,1,...,n-1$) combined with inversion.
Since the invariant space of $\widetilde{C}_{n,p}P$
is just TRIMs on the rotation axis of $\widetilde{C}_{n,p}$, 
one can expect at most Dirac points on the rotation axis, which has already been extensively studied~\cite{Yang_classification1,Yang_classification2}. 
On the other hand, in the case of $\widetilde{C}_{4,p=1,3}^{\parallel}$ ($\widetilde{C}_{6,p=1,3,5}^{\parallel}$) symmetry which can generate two-fold screw rotation $\left[\widetilde{C}_{4,p}^{\parallel}\right]^{2}$ ($\left[\widetilde{C}_{6,p}^{\parallel}\right]^{3}$) and the associated off-centered mirror $\widetilde{M}^{\perp}$, one may consider additional constraints on Dirac line nodes imposed
by $\widetilde{C}_{4,p}^{\parallel}$ ($\widetilde{C}_{6,p}^{\parallel}$).
It is straightforward to show that $\widetilde{C}_{4,p=1,3}^{\parallel}$ or $\widetilde{C}_{6,p=1,3,5}^{\parallel}$ does not commute with $\widetilde{M}^{\perp}$
in the mirror plane on the zone boundary where Dirac line nodes are expected,
thus the screw rotation does not affect the distribution
of $\widetilde{M}^{\perp}$ eigenvalues.
One exception is the case with $\widetilde{C}_{6,3}^{\parallel}$ symmetry.
Since $\left[\widetilde{C}_{6,3}^{\parallel}\right]^{2}=\{C_{3}|\bm{0}\}$
commutes with $\widetilde{M}^{\perp}$, the distribution of $\widetilde{M}^{\perp}$
also satisfies three-fold rotation symmetry, which constrains
the number of open line nodes to be a multiple of 3.

It is worth to note that the presence of inversion symmetry is not necessary to
have off-centered symmetries.
Off-centered symmetries can exist, in general,
as long as a nonsymmorphic crystal with screw/glide symmetries contain an
additional point group symmetry such as mirror symmetry
whose reference point does not coincide with that
of screw/glide symmetries.
In noncentrosymmetric systems, stable Weyl point/line nodes with two-fold degeneracy can be protected by off-centered symmetry when two bands with different eigenvalues cross
in the relevant invariant space.

Up to now, we have considered $\bm{t}_{\perp}$ and $\bm{t}_{\parallel}$ separately.
However, in many nonsymmorphic crystals, $\bm{t}_{\perp}$ and $\bm{t}_{\parallel}$
coexist, which gives rise to {\it off-centered screw/glide symmetries}.
Interestingly, the off-centered two-fold screw/glide symmetry can protect a single point/line node with four-fold degeneracy in an invariant space. For instance, let us consider a system corresponding to the space group No.14 containing
$T$, $\{P|\bm{0}\}$, and an off-centered glide mirror $\widetilde{M}_{z}^{\parallel,\perp}=\{M_{z}|\frac{1}{2}\hat{x}+\frac{1}{2}\hat{z}\}$ with
$\bm{t}_{\perp}=\frac{1}{2}\hat{z}$ and $\bm{t}_{\parallel}=\frac{1}{2}\hat{x}$. On the $k_{z}=\pi$ plane,
the Kramers degenerate $\widetilde{M}_{z}^{\parallel,\perp}$ eigenstates at each momentum have the same $\widetilde{M}_{z}^{\parallel,\perp}$ eigenvalues
$\pm ie^{i\frac{1}{2}k_{x}}$.
However, due to their momentum dependence,
when $k_{x}$ is shifted by $2\pi$, two different $\widetilde{M}_{z}^{\parallel,\perp}$ eigensectors should be interchanged.
This naturally gives rise to an open line node connecting two TRIMs at $\bm{k}=(0,0,\pi)$
and $(0,\pi,\pi)$ as shown in Fig.~\ref{fig:offcentered_glide} (d-f).
Contrary to the semimetal protected by $\widetilde{M}_{x}^{\perp}$
with an even number of nodal lines,
the semimetal protected by the off-centered glide mirror $\widetilde{M}_{z}^{\parallel,\perp}$
has a single nodal line. This is because, due to the $k_{x}$ dependence of $\widetilde{M}_{z}^{\parallel,\perp}$
eigenvalues, it is possible to get around the doubling in the number of line nodes~\cite{Yang_classification2}. (See Methods.)
Repeating a similar analysis, one can easily see that
an off-centered two-fold screw rotation, $\widetilde{C}_{2z}^{\parallel,\perp}=\widetilde{M}_{z}^{\parallel,\perp}P=\{C_{2z}|\frac{1}{2}\hat{x}+\frac{1}{2}\hat{z}\}$, has momentum-dependent eigenvalues, which can give rise to a semimetal with a single Dirac point
on each rotation axis. (See Fig.~\ref{fig:offcentered_glide} (a-c).) 

Let us note that a line node semimetal protected by an off-centered
glide mirror is already discussed in Ref.~\onlinecite{linenode_new_Fu}. Although the key role
of an off-centered glide mirror on the protection of a nodal line with four-fold
degeneracy is correctly described in this work, a line node predicted in Ref.\onlinecite{linenode_new_Fu} forms a closed loop, which is not consistent with our theory. We believe that
a nodal line protected by a single off-centered glide mirror should have an open shape. Correct description of the shape of nodal lines is important to resolve
the controversies related with the mechanism protecting the circular Dirac line node in SrIrO$_{3}$. To explain the origin of the Dirac line node
in SrIrO$_{3}$, several different ideas are proposed including off-centered glide mirror symmetry~\cite{linenode_new_Fu}, simultaneous presence of mirror and chiral symmetries~\cite{linenode_HY}, the presence of multiple nonsymmorphic symmetries~\cite{linenode_new_HY}. According to our theoretical analysis,
we believe that the presence of a single nonsymmorphic symmetry can protect
only line nodes with open shape. We think that the presence of multiple nonsymmorphic symmetries is necessary to describe the circular nodal line in SrIrO$_3$ as proposed in Ref.~\onlinecite{linenode_new_HY}.

BaTaS$_3$ is another material, which has stable Dirac line nodes with fourfold degeneracy in the presence of spin-orbit coupling. Interestingly, according to a recent first-principles calculation~\cite{linenode_new_Weng}, it is found that the nodal lines in this material have open shape,
which is consistent with our model calculation shown in Fig.~\ref{fig:typeII_mirror} (b).
In this system, due to the additional mirror symmetry whose invariant plane
is orthogonal to that of the off-centered mirror symmetry, the nodal lines have
open shape similar to the case shown in Fig.~\ref{fig:multiple_mirror} (a).
Though the role of the additional mirror symmetry is emphasized in Ref.~\onlinecite{linenode_new_Weng}, based on our theoretical consideration,
we think that the presence of $P$, $T$, and an off-centered mirror is sufficient for the protection of the Dirac line node itself.

To sum up, we propose a general theoretical framework to understand
a class of 3D semimetals with Dirac line/point nodes with four-fold degeneracy,
which are stable in the presence of strong spin-orbit coupling. 
We have identified
the presence of {\it off-centered crystalline symmetries}
as a mechanism for the protection of the point/line nodes.
If the crystalline symmetries relevant to the protection of each nodal semimetal
is partially lifted by applying external magnetic field
or doping magnetic ions,
one can observe a significant change of Fermi surface topology,
involved with the emerging topological semimetals with Weyl point nodes or Weyl line nodes,
or even a gapped insulator. Such a tunability of the Fermi surface topology under magnetic field
can provide a promising venue for various intriguing topological magneto-transport phenomena.
Moreover, since both
weak disorder (Coulomb potential) is irrelevant (marginally irrelevant) perturbation,
the prediction based on noninteracting semimetals is perturbatively valid 
even in disordered (interacting) systems~\cite{Sudip, Yejin}.
To understand the role of strong electron correlation and disorder,
and, in particular, the combined effect of them are important
issues to be studied in future research.

\subsection*{Methods}
{\bf Lattice model.}
In the momentum space, the tight-binding Hamiltonian in Eq.~(\ref{eqn:model_Rspace}) can be written as
$\hat{H}^{(0)}=\sum_{\bm{k}}\hat{c}^{\dag}(\bm{k})H^{(0)}(\bm{k})\hat{c}(\bm{k})$ with
\begin{align}
H^{(0)}(\bm{k})=&-\left[\left(t_{1}+t_{2}\cos k_{z}\right)\tau_{x}+t_{2}\sin k_{z}\tau_{y}\right]\cos\frac{k_{x}}{2}\cos\frac{k_{y}}{2}
-t_{3}(\cos k_{x}+\cos k_{y})-t_{4}\cos k_{z}
\nonumber\\
&+(\lambda_{1}-\lambda_{2}\cos k_{z})(\sin k_{x}\sigma_{y}-\sin k_{y}\sigma_{x})\tau_{z},
\end{align}
where the Pauli matrices $\tau_{x,y,z}$ denote the sublattice degrees of freedom.
$\hat{H}^{(0)}$ is invariant not only under the time-reversal symmetry $T$ and
the inversion symmetry $P$ but also under the off-centered symmetries
$\widetilde{C}^{\perp}_{2z}=\{C_{2z}|\frac{1}{2}\hat{x}+\frac{1}{2}\hat{y}\}$,
$\widetilde{M}^{\perp}_{x}=\{M_{x}|\frac{1}{2}\hat{x}\}$, and
$\widetilde{M}^{\perp}_{y}=\{M_{y}|\frac{1}{2}\hat{y}\}$.
At the $\Gamma$ point the symmetry operators are written as
$T=i\sigma_y\mathcal{K}$, $P=\tau_x$,
$\widetilde{C}^\perp_{2z}=i\sigma_z$,
$\widetilde{M}^\perp_x=i\sigma_x$, and
$\widetilde{M}^\perp_y=i\sigma_y$,
where $\mathcal{K}$ is complex conjugation operator.

To understand the role of each off-centered symmetry,
we distort the lattice in two different ways.
First, we shift the position of the B site in a unit cell in the $y$ direction,
which makes $\bm{r}_{AB}=(\frac{1}{2},\delta_{y}\neq\frac{1}{2},\delta_{z})$.
This distortion breaks $\widetilde{C}^{\perp}_{2z}$ and $\widetilde{M}^{\perp}_{y}$ symmetries,
and generates the following additional terms in the Hamiltonian
\begin{align}\label{eqn:Hpert1}
\delta H^{(1)}(\bm{k})=&
\left[\left(t'_{1}+t'_{2}\cos k_{z}\right)\tau_{y}-t'_{2}\sin k_{z}\tau_{x}\right]\cos\frac{k_{x}}{2}\sin\frac{k_{y}}{2}
\nonumber\\
&+(\lambda'_{1}-\lambda'_{2}\cos k_{z})\sin k_{x}\sigma_{z}\tau_{z}
-[\lambda'_{2}(\cos k_{x}+ \cos k_{y})+\lambda_{3}]\sin k_{z}\sigma_{x}\tau_{z}.
\end{align}
The full Hamiltonian $H^{(0)}(\bm{k})+\delta H^{(1)}(\bm{k})$ supports
nodal lines in the $k_{x}=\pi$ plane protected by
the $\widetilde{M}^{\perp}_{x}$ symmetry.
The $8\times8$ Hamiltonian with the band structure shown
in Fig.~\ref{fig:linenode_doubling}(b) is given by
$[H^{(0)}(\bm{k})+\delta H^{(1)}(\bm{k})]\upsilon_0
+\lambda_4\sigma_x\upsilon_y$,
where $\upsilon_0$ is a $2\times2$ unit matrix and $\upsilon_{x,y,z}$
are Pauli matrices in the new grading.

The second distortion is achieved by deforming the lattice along
the [110] direction, which breaks $\widetilde{M}^{\perp}_{x}$,
$\widetilde{M}^{\perp}_{y}$ symmetries
and generates the following term,
\begin{align}
\delta H^{(2)}(\bm{k})=&
-\left[\left(t'_{1}+t'_{2}\cos k_{z}\right)\tau_{x}+t'_{2}\sin k_{z}\tau_{y}\right]\sin\frac{k_{x}}{2}\sin\frac{k_{y}}{2}
\nonumber\\
&-(\lambda'_{1}+\lambda'_{2}\cos k_{z})(\sin k_{x}\sigma_{x}-\sin k_{y}\sigma_{y})\tau_{z}
+\lambda'_{2}\sin k_{z}(\cos k_{x}-\cos k_{y})\sigma_{z}\tau_{z}.
\end{align}
The full Hamiltonian $H^{(0)}(\bm{k})+\delta H^{(2)}(\bm{k})$ supports
nodal points protected by the $\widetilde{C}^{\perp}_{2z}$ symmetry.
The $8\times8$ Hamiltonian with the band structure shown in
Fig.~\ref{fig:pointnode_doubling}(b) is given by
$[H^{(0)}(\bm{k})+\delta H^{(2)}(\bm{k})]\upsilon_0
+\lambda_5\sigma_z\upsilon_y+\epsilon\upsilon_z$.

We have used the following parameters in the numerical calculations:
$t_1 = 1$,
$t_2 = 0.15$,
$t_3 = 0.3$,
$t_4 = 0.1$,
$\lambda_1 = 1.5$,
$\lambda_2 = 0.4$,
$t_1' = 0.5$,
$t_2' = 0.15$,
$\lambda_1' = 0.7$,
$\lambda_2' = 0.2$,
$\lambda_3 = 0.2$,
$\lambda_4 = 0.12$,
$\lambda_5 = 0.05$,
and
$\epsilon = 0.05$.

{\bf Topological charge.}
The topological charges of a point node protected by $\widetilde{C}^{\perp}_{2z}$
and a line node protected by $\widetilde{M}^{\perp}_{x}$ can be determined as follows.
First, for a point node, a zero-dimensional topological invariant $Q$ is defined as
\begin{align}
Q=\frac{1}{8}\left[N(k_{N})-N(k_{S})\right]\in \mathbb{Z},
\end{align}
where $k_{N}$ ($k_{S}$) is the $k_{z}$ momentum on the $\widetilde{C}^{\perp}_{2z}$ invariant axis
slightly above (below) the Dirac point, and $N(k_{z})$ is given by
\begin{align}
N(k_{z})=N_{+}(k_{z})-N_{-}(k_{z}),
\qquad
N_{\pm}(k_{z})=N^{c}_{\pm}(k_{z})-N^{v}_{\pm}(k_{z}),
\label{N(k_z)}
\end{align}
where $N^{c}_{\pm}(k_{z})$ and $N^{v}_{\pm}(k_{z})$
denote the numbers of the conduction and valence bands
with the $\widetilde{C}^{\perp}_{2z}$ eigenvalues of $\pm i$
at the momentum $k_{z}$.
Let us note that the sum $N_{+}(k_{z})+N_{-}(k_{z})$
is constant on the $\widetilde{C}^{\perp}_{2z}$ invariant axis
whereas the difference $N(k_{z})=N_{+}(k_{z})-N_{-}(k_{z})$
can take an integer value, thus provides a topological index characterizing
a Dirac point, which jumps across a Dirac point.

\begin{figure*}[t]
\centering
\includegraphics[width=16 cm]{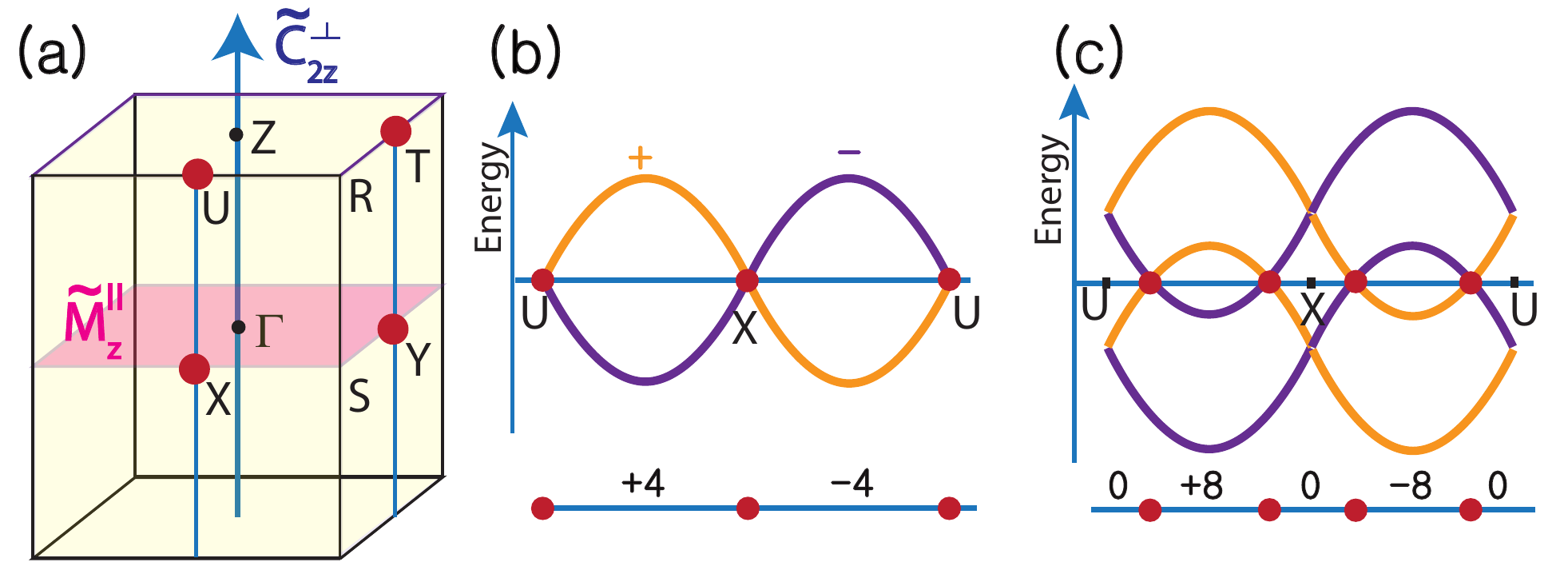}
\caption{
{\bf 3D Dirac points protected by an off-centered two-fold rotation $\widetilde{C}^{\perp}_{2z}$.}
({\bf a})
A schematic figure describing the distribution of 3D Dirac points in momentum space.
The location of four 3D Dirac points is marked in red dots.
The bold blue arrow indicates the axis for $\widetilde{C}^{\perp}_{2z}$ symmetry.
The pink square indicates the plane of the glide mirror symmetry $\widetilde{M}^{\parallel}_{z}$,
which is dual to $\widetilde{C}^{\perp}_{2z}$.
({\bf b})
The band structure along the $U-X-U$ line on which $\{P,\widetilde{C}^{\perp}_{2z}\}=0$.
A pair of degenerate bands (a doublet pair) form a 3D Dirac point at each time-reversal invariant momentum (TRIM).
The corresponding distribution of topological charges is shown in the bottom.
({\bf c})
The band structure when two doublet pairs cross on the $U-X-U$ line.
There are in total $4n$ ($n$ is an integer) 3D Dirac points, which are all away from TRIMs.
The corresponding distribution of the topological charges, whose definition is given in Methods, 
is shown in the bottom.
} \label{fig:typeII_rotation}
\end{figure*}

\begin{figure*}[t]
\centering
\includegraphics[width=16 cm]{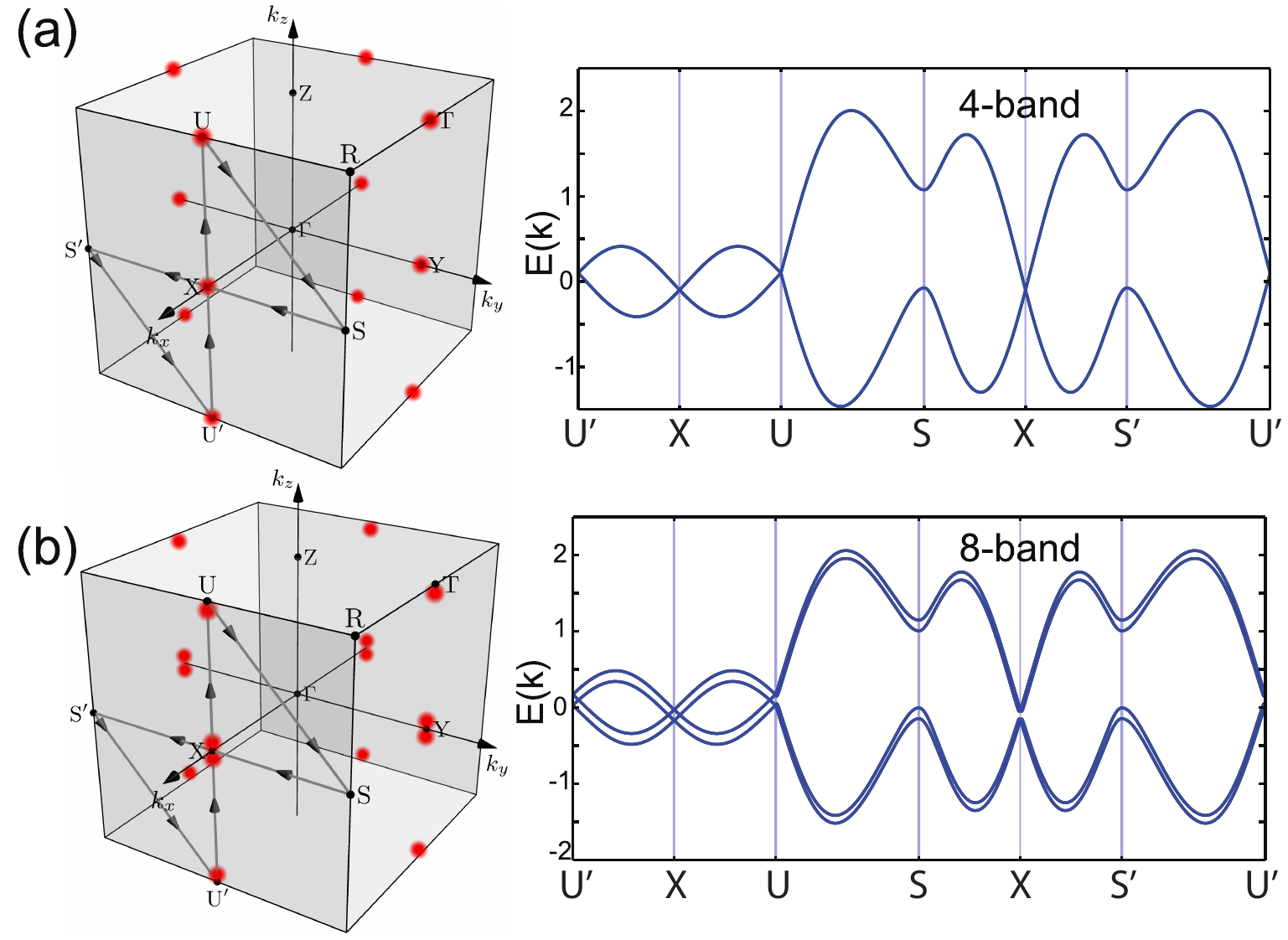}
\caption{
{\bf Band structure of a semimetal with Dirac point nodes protected by 
$\widetilde{C}^{\perp}_{2z}=\{C_{2z}|\frac{1}{2}\hat{x}+\frac{1}{2}\hat{y}\}$ symmetry.}
({\bf a})
From 4-band lattice models with $P$, $T$, and $\widetilde{C}^{\perp}_{2z}$ symmetries.
Doublet pairs on the $\bm{k}=(\pi,0,k_{z})$ and $\bm{k}=(0,\pi,k_{z})$ 
lines with $k_{z}\in[-\pi,\pi]$, form Dirac point nodes at every TRIM. 
A bold red dot on the left marks the points where two bands stick together.
({\bf b})
Similar plots from a 8-band lattice model.
Crossing points between two doublet pairs form four Dirac points away
from TRIM in both $\bm{k}=(\pi,0,k_{z})$ and $\bm{k}=(0,\pi,k_{z})$ 
lines with $k_{z}\in[-\pi,\pi]$, which correspond to those marked
by red circles in Fig.~\ref{fig:typeII_rotation}(c).
In the figures, $U'$ and $S'$ indicate the momenta equivalent to $U$ and $S$, respectively. 
} \label{fig:pointnode_doubling}
\end{figure*}

There are several constraints on $N(k_{z})$ imposed by the time-reversal
and the inversion symmetries.
Firstly, the anti-unitary time-reversal symmetry,
which commutes with the $\widetilde{C}^{\perp}_{2z}$ requires
\begin{align}
N_{\pm}(k_{z})=N_{\mp}(-k_{z}),
\end{align}
thus
\begin{align}\label{eqn:point_constraint_T}
N(k_{z})=-N(-k_{z}).
\end{align}
On the other hand, the inversion symmetry anti-commuting
with $\widetilde{C}'_{2z}$ requires
\begin{align}\label{eqn:contraint_Npm}
N_{\pm}(k_{z})=N_{\mp}(-k_{z}),
\end{align}
thus again
\begin{align}\label{eqn:point_constraint_P}
N(k_{z})=-N(-k_{z}).
\end{align}
The constraint in Eq.~(\ref{eqn:point_constraint_T}) and
Eq.~(\ref{eqn:point_constraint_P}) naturally leads
to the band structure and the corresponding distribution of
$N(k_z)$ shown in
Fig.~\ref{fig:typeII_rotation}.

In the case of a line node protected by $\widetilde{M}^{\perp}_{x}$,
a zero-dimensional topological invariant $Q'$ can be defined as follows.
For a given gapless point $\bm{k}$ on a line node in the $k_{x}=\pi$ plane,
one can find two points $\bm{k}_{N}$ and $\bm{k}_{S}$
in a way that the line connecting them is normal to the tangential vector
at $\bm{k}$.
Then the topological charge of a line node can be defined in the exactly
the same way as in the case of the point node.
Namely, a zero-dimensional topological invariant is defined as
\begin{align}
Q'=\frac{1}{8}\left[N(\bm{k}_{N})-N(\bm{k}_{S})\right]
\in \mathbb{Z}
\end{align}
where the definition of $N(\pi,k_{y},k_{z})$ is exactly the same as
Eq.\ (\ref{N(k_z)}).
The topological invariant $Q'$ measures
the change in $N(\bm{k})$
across a line node
in the $k_{x}=\pi$ plane.

\begin{figure*}[t]
\centering
\includegraphics[width=16 cm]{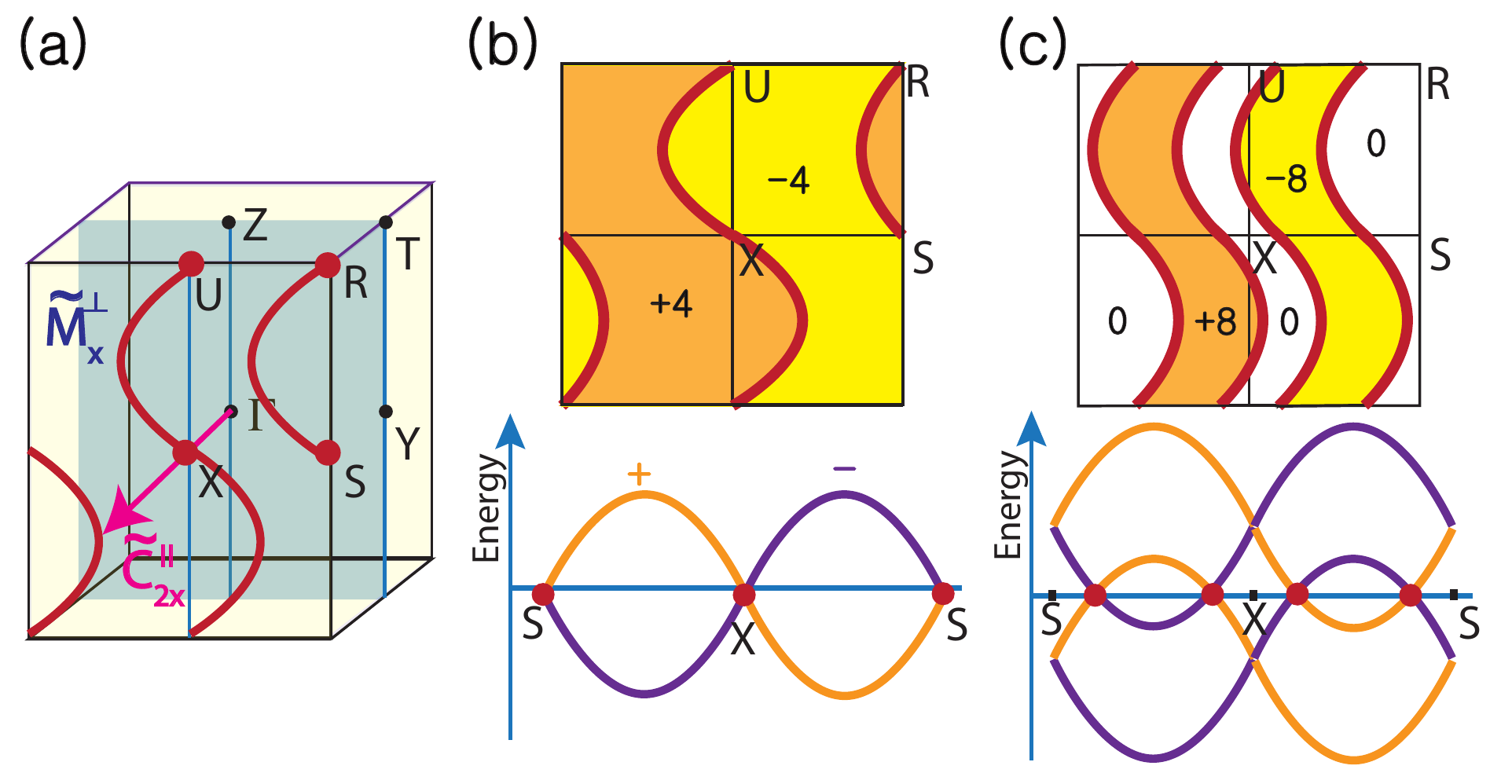}
\caption{
{\bf 3D Dirac lines protected by an off-centered mirror symmetry $\widetilde{M}^{\perp}_{x}$.}
({\bf a})
A schematic figure describing the distribution of nodal lines in momentum space.
The location of two line nodes in the $k_{x}=\pi$ plane is marked in red color.
The blue square indicates the plane of the off-centered mirror symmetry $\widetilde{M}^{\perp}_{x}$.
The bold pink arrow indicates the axis for the screw rotation $\widetilde{C}^{\parallel}_{2x}$ symmetry,
which is dual to $\widetilde{M}^{\perp}_{x}$ symmetry.
({\bf b})
Distribution of topological charges in the $k_{x}=\pi$ plane in which $\{P,\widetilde{M}^{\perp}_{x}\}=0$
is satisfied.
The corresponding band structure along the $S-X-S$ line is shown in the bottom.
The doublet pair are degenerate at each TRIM, which is a part of line nodes
in the $k_{x}=\pi$ plane.
({\bf c})
Distribution of topological charges, which are defined in Methods, 
when two doublet pairs cross in the $k_{x}=\pi$ plane.
There are in total $4n$ ($n$ is an integer) nodal lines, which are away from TRIM.
The corresponding band structure along the $S-X-S$ line is shown in the bottom.
} \label{fig:typeII_mirror}
\end{figure*}

\begin{figure*}[t]
\centering
\includegraphics[width=16 cm]{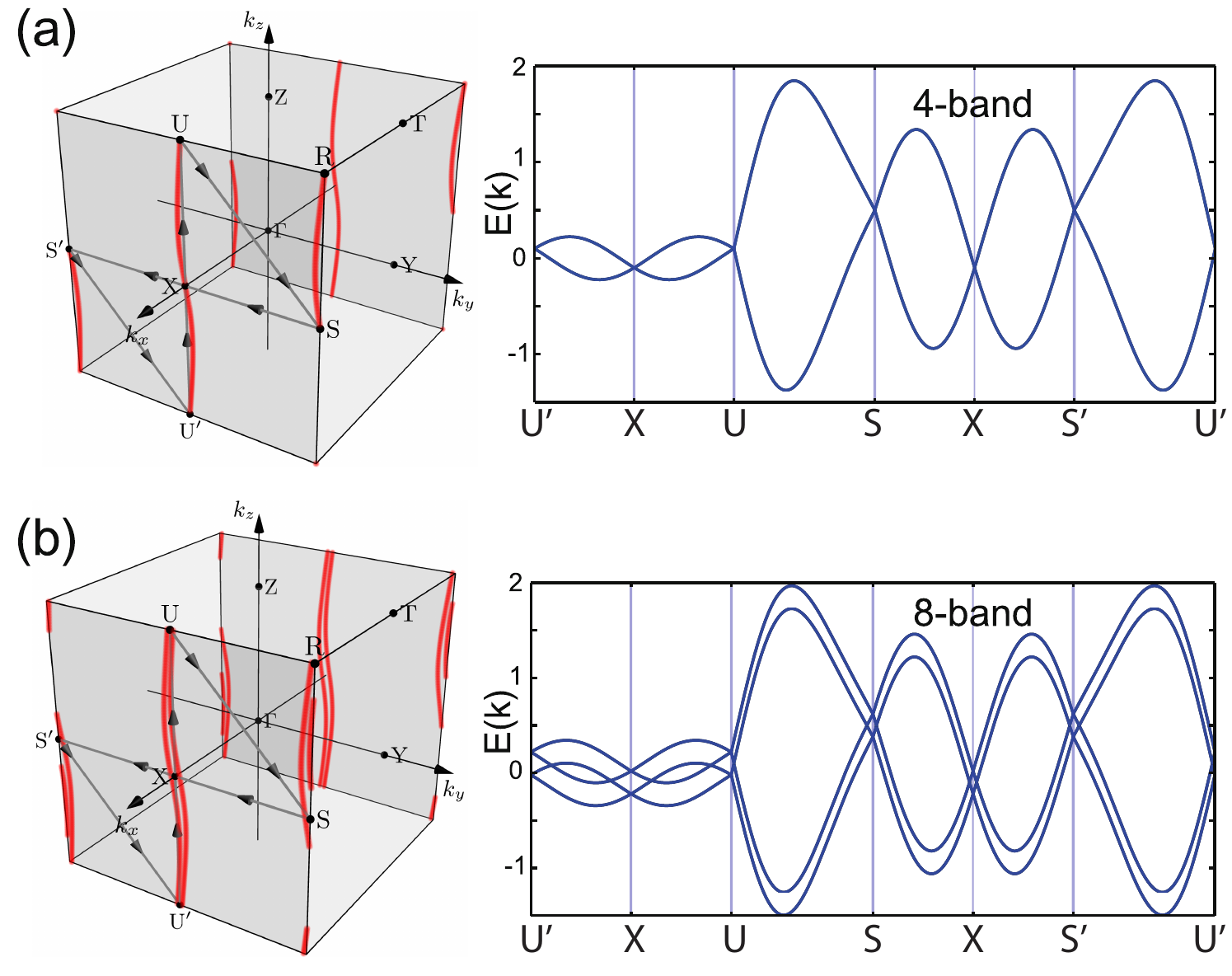}
\caption{
{\bf Band structure of a semimetal with Dirac line nodes protected by 
$\widetilde{M}^{\perp}_{x}=\{M_{x}|\frac{1}{2}\hat{x}\}$ symmetry.}
({\bf a})
From 4-band lattice models with $P$, $T$, and $\widetilde{M}^{\perp}_{x}$ symmetries.
Here eigenstates are doubly degenerate at each momentum.
Doublet pairs on the $k_{x}=\pi$ plane form two Dirac line nodes, each passes two TRIMs. 
The bold red lines in the left panel mark the points 
in the BZ  where two bands stick together.
({\bf b})
Similar plots from a 8-band lattice model.
Crossing points between two doublet pairs form four Dirac line nodes on the $k_{x}=\pi$ plane, which correspond to those marked
by red circles in Fig.~\ref{fig:typeII_mirror}(c).
In the figures, $U'$ and $S'$ indicate the momenta equivalent to $U$ and $S$, respectively. 
} \label{fig:linenode_doubling}
\end{figure*}

There are several constraints on $N(\pi,k_{y},k_{z})$ imposed by the time-reversal
and the inversion symmetries.
Firstly, the anti-unitary
time-reversal symmetry, which commutes with the $\widetilde{M}'_{x}$ requires
\begin{align}
N_{\pm}(\pi,k_{y},k_{z})=N_{\mp}(\pi,-k_{y},-k_{z}),
\end{align}
thus
\begin{align}\label{eqn:line_constraint_T}
N(\pi,k_{y},k_{z})=-N(\pi,-k_{y},-k_{z}).
\end{align}
On the other hand, the inversion symmetry anti-commuting with $\widetilde{M}'_{x}$ requires
\begin{align}\label{eqn:contraint_Npm2}
N_{\pm}(\pi,k_{y},k_{z})=N_{\mp}(\pi,-k_{y},-k_{z}),
\end{align}
thus again
\begin{align}\label{eqn:line_constraint_P}
N(\pi,k_{y},k_{z})=-N(\pi,-k_{y}-k_{z}).
\end{align}
The constraint in Eq.~(\ref{eqn:line_constraint_T}) and
Eq.~(\ref{eqn:line_constraint_P}) naturally leads
to the band structure and the corresponding distribution of
$N(\pi,k_y,k_z)$
shown in Fig.~\ref{fig:typeII_mirror}.

It is worth to note that both time-reversal symmetry and inversion symmetry put the same
constraint on the topological invariants as shown in Eq.~(\ref{eqn:point_constraint_T}) and Eq.~(\ref{eqn:point_constraint_P}),
and also in Eq.~(\ref{eqn:line_constraint_T}) and Eq.~(\ref{eqn:line_constraint_P}).
This means that as long as 
either $P$ or $T$ is preserved, the eigenstates always form pairs carrying quantized eigenvalues
of $\widetilde{C}^{\perp}_{2z}$ or $\widetilde{M}^{\perp}_{x}$ in the relevant invariant space. 
Since an eigenstate at a generic momentum is non-degenerate when either $T$ or $P$ is broken,
a pair is composed of non-degenerate states in this case.
Hence whenever there is crossing of states having different eigenvalues of the relevant off-centered symmetries,
Weyl point/line nodes with two-fold degeneracy can be created.

{\bf Clifford algebras and stability of nodal points and nodal lines.}
Here we use Clifford algebras~\cite{Kitaev,Morimoto1,Morimoto2,ZDWang1,ZDWang2,Sato1,Sato2} to show the existence
of stable Dirac points under $\widetilde{C}^{\perp}_{2z}$
and stable Dirac line nodes under $\widetilde{M}^{\perp}_{x}$,
and determine the relevant topological charges.

First, we show that stable Dirac points protected by
$\widetilde{C}^{\perp}_{2z}$ can exist at
the four TRIMs at $\bm{k}=(\pi,0,0)$,
$(\pi,0,\pi)$, $(0,\pi,0)$, and $(0,\pi,\pi)$,
where $\{P,\widetilde{C}^{\perp}_{2z}\}=0$ and
$(\widetilde{C}^{\perp}_{2z})^2=-1$.
Suppose that the effective Dirac Hamiltonian around a TRIM has
a Dirac mass term $\gamma_0$,
\begin{equation}
H=q_x\gamma_x+q_y\gamma_y+q_z\gamma_z+m\gamma_0,
\label{3D Dirac}
\end{equation}
where $\bm{q}$ is the momentum measured from the TRIM,
$m$ is the Dirac mass,
and the gamma matrices mutually anticommute.
The velocity is set equal to unity.
We are going to see that the Dirac mass term $m\gamma_0$ is
not allowed by symmetries.

Since $T$ and $P$ changes $\bm{q}\to-\bm{q}$, the invariance of the
Dirac Hamiltonian requires the gamma matrices to obey
\begin{align}
&
\{T,\gamma_x\}=\{T,\gamma_y\}=\{T,\gamma_z\}=[T,\gamma_0]=0,
\label{gamma T}\\
&
\{P,\gamma_x\}=\{P,\gamma_y\}=\{P,\gamma_z\}=[P,\gamma_0]=0.
\label{gamma P}
\end{align}
The symmetry operators $T$, $P$, and $\widetilde{C}^{\perp}_{2z}$
satisfy the following relations:
\begin{align}
T^2=-1,
\qquad
P^2=1,
\qquad
(\widetilde{C}^{\perp}_{2z})^2=-1,
\\
[T,P]=[T,\widetilde{C}^{\perp}_{2z}]=\{P,\widetilde{C}^{\perp}_{2z}\}=0.
\end{align}
Since $\widetilde{C}^{\perp}_{2z}$ changes $q_{x,y}\to-q_{x,y}$,
the invariance of the Hamiltonian under $\widetilde{C}^{\perp}_{2z}$
leads to
the relations
\begin{equation}
\{\widetilde{C}^{\perp}_{2z},\gamma_x\}=
\{\widetilde{C}^{\perp}_{2z},\gamma_y\}=
[\widetilde{C}^{\perp}_{2z},\gamma_z]=
[\widetilde{C}^{\perp}_{2z},\gamma_0]=0.
\end{equation}

The Clifford algebra is constructed from the gamma matrices and
the symmetry operators,
\begin{equation}
Cl_{4,4}=
\{T,JT,J\gamma_0,\widetilde{C}^{\perp}_{2z}P\gamma_z;
\gamma_x,\gamma_y,\gamma_z,P\gamma_x\gamma_y\gamma_z\},
\end{equation}
where $J$ represents the imaginary unit ``$i$''
satisfying $J^2=-1$ and $\{T,J\}=0$.
Here we have used the notation for the Clifford algebra
$Cl_{p,q}=\{e_1,\ldots,e_p;e_{p+1},\ldots,e_{p+q}\}$, where
the generators $e_j$ mutually anticommute and satisfy
$e_j^2=-1$ for $j=1,\ldots,p$ and $e_j^2=+1$ for $j=p+1,\ldots,p+q$.
The existence/absence condition of the Dirac mass $m\gamma_0$ is
determined by the extension problem\cite{Morimoto2}
\begin{equation}
Cl_{2,4}\to Cl_{3,4}.
\end{equation}
The relevant classifying space is $R_{2-4+2}=R_0$, and
$\pi_0(R_0)=\mathbb{Z}$.
This implies that no Dirac mass is available,
and a Dirac point has a $\mathbb{Z}$ topological charge.

We have seen in Fig.~\ref{fig:pointnode_doubling}(b) that
Dirac point nodes are shifted from TRIMs in the 8-band model.
Their stability can be understood with Clifford algebra as follows.
We consider the Dirac Hamiltonian in Eq.~(\ref{3D Dirac}) with
$(q_x,q_y,q_z)$ measured from a Dirac point away from TRIMs.
In this case $T$ and $P$ are not independent symmetries,
and the product $TP$ is the symmetry operator to be considered.
It satisfies the relations
\begin{equation}
(TP)^2=-1,
\qquad
[TP,\gamma_x]=[TP,\gamma_y]=[TP,\gamma_z]=[TP,\gamma_0]=0,
\qquad
\{TP,\widetilde{C}^\perp_{2z}\}=0.
\end{equation}
From he gamma matrices $\gamma_\mu$ and the symmetry operators,
we can construct the Clifford algebra
\begin{equation}
Cl_{0,4}\otimes Cl_{2,1}=
\{;\gamma_x,\gamma_y,\gamma_z,\gamma_0\}\otimes
\{TP,JTP;\gamma_x\gamma_y\widetilde{C}^\perp_{2z}\}.
\end{equation}
The existence condition of the Dirac mass term $m\gamma_0$ is
determined from the extension problem
\begin{equation}
Cl_{0,2}\otimes Cl_{2,1}\to Cl_{0,3}\otimes Cl_{2,1},
\end{equation}
which is equivalent to the extension of complex Clifford algebra
$Cl_4\to Cl_5$, for which the relevant classifying space is $C_4=C_0$.
Since $\pi_0(C_0)=\mathbb{Z}$, no Dirac mass term is available,
and a Dirac point has an integer topological index.

Next, we prove the stability of the line nodes protected
by $\widetilde{M}^{\perp}_{x}$ in the $k_{x}=\pi$ plane
where $\{P,\widetilde{M}^{\perp}_{x}\}=0$
and $(\widetilde{M}^{\perp}_{x})^2=-1$.
We prove the stability of line nodes in two steps .
First, we show that two bands touch at TRIMs with $k_x=\pi$.
We consider the Dirac Hamiltonian
\begin{equation}
H=q_x\gamma_x+q_y\gamma_y+m\gamma_0,
\label{Dirac Hamiltonian 2}
\end{equation}
where $(q_x,q_y)$ is a momentum measured from a TRIM with $k_z=0$ or $\pi$.
The gamma matrices satisfy the relations in Eqs.\ (\ref{gamma T})
and (\ref{gamma P}).
The symmetry operators satisfy
\begin{align}
T^2=-1,
\qquad
P^2=1,
\qquad
(\widetilde{M}^{\perp}_{x})^2=-1,\qquad
[T,P]=[T,\widetilde{M}^{\perp}_{x}]=\{P,\widetilde{M}^{\perp}_{x}\}=0.
\end{align}
Since $\widetilde{M}^{\perp}_{x}$ changes the momentum
$\bm{q}=(q_x,q_y,q_z)\to(-q_x,q_y,q_z)$,
the commutation relations between gamma matrices and $\widetilde{M}^{\perp}_{x}$ are
given by
\begin{equation}
\{\widetilde{M}^{\perp}_{x},\gamma_x\}=[\widetilde{M}^{\perp}_{x},\gamma_y]
=[\widetilde{M}^{\perp}_{x},\gamma_0]=0.
\label{C_2x,gamma}
\end{equation}
Around TRIMs, $T$ and $P$ can be treated as symmetry operators.
We can construct the Clifford algebra
\begin{equation}
Cl_{4,2}\otimes Cl_{1,0}
=\{T,TJ,J\gamma_0,\widetilde{M}^{\perp}_{x}P\gamma_y;\gamma_x,\gamma_y\}
\otimes\{P\gamma_x\gamma_y;\}.
\end{equation}
The existence condition of the Dirac mass term $m\gamma_0$ is
determined by the extension problem
\begin{equation}
Cl_{2,2}\otimes Cl_{1,0}\to Cl_{3,2}\otimes Cl_{1,0},
\end{equation}
or equivalently, $Cl_4\to Cl_5$.
The relevant classifying space is $C_4=C_0$, and $\pi_0(C_0)=\mathbb{Z}$.
Thus, no Dirac mass term is available.
This means that the energy levels at these TRIMs are four-fold degenerate.

Next we fix $k_z$ to be constant different from $0$ and $\pi$.
We consider the Dirac Hamiltonian (\ref{Dirac Hamiltonian 2}),
where $q_x$ and $q_y$ are now understood to be momentum measured from
a line node on the constant $k_z$ plane.
Away from TRIMs, $T$ and $P$ are not independent symmetries,
and instead the product $TP$ is the symmetry operator to be considered.
It obeys the relations
\begin{equation}
(TP)^2=-1,
\qquad
[TP,\gamma_x]=[TP,\gamma_y]=[TP,\gamma_0]=0,\qquad \{TP,\widetilde{M}^{\perp}_{x}\}=0.
\end{equation}
Let's consider if the Dirac mass term $m\gamma_0$ is allowed in
the Dirac Hamiltonian (\ref{Dirac Hamiltonian 2})
under the symmetries $TP$ and $\widetilde{M}^{\perp}_{x}$.
The Clifford algebra is constructed from the symmetry operators
and the gamma matrices:
\begin{equation}
Cl_{1,3} \otimes Cl_{2,0}
 = \{J\gamma_x\widetilde{M}^{\perp}_{x}; \gamma_x, \gamma_y, \gamma_0\} \otimes \{TP, JTP; \}.
\end{equation}
The extension problem
$Cl_{1,1}\otimes Cl_{2,0}\to Cl_{1,2}\otimes Cl_{2,0}$
is equivalent to
$Cl_{3,1} \to Cl_{4,1}$, which leads to the classifying space
$R_{3+2-1}=R_4$;
$\pi_0(R_4) = \mathbb{Z}$.
This means that a line node
is topologically stable on the $k_x=\pi$ plane.

{\bf Low energy $\bm{k}\cdot\bm{p}$ Hamiltonian analysis.}
Here we construct the low energy effective $4\times 4$ Hamiltonian
near the momentum $\bm{k}=(\pi,0,0)$ to understand
the nature of nodal points/lines depending on the symmetry of the system.
We take the following representation of symmetry operators:
\begin{equation}
T=i\sigma_y\tau_z\mathcal{K},
\qquad
P=-\tau_y,
\qquad
\widetilde{C}^{\perp}_{2z}=i\sigma_z\tau_z,
\qquad
\widetilde{M}^{\perp}_{x}=i\sigma_x\tau_z,
\qquad
\widetilde{M}^{\perp}_{y}=i\sigma_y.
\end{equation}
The effective Hamiltonian can be obtained by collecting all symmetry-allowed operators up to linear order in
$\bm{q}=\bm{k}-(\pi,0,0)$.
In the following analysis we omit terms proportional to
the unit $4\times4$ matrix for simplicity.

Firstly, the effective Hamiltonian
invariant under $T$, $P$, and $\widetilde{C}^{\perp}_{2z}$
has the form
\begin{equation}
H_0=
q_x(v_{1x}\tau_x+v_{2x}\sigma_x\tau_z+v_{3x}\sigma_y\tau_z)
+q_y(v_{1y}\tau_x+v_{2y}\sigma_x\tau_z+v_{3y}\sigma_y\tau_z)
+v_zq_z\sigma_z\tau_z
\end{equation}
with energy eigenvalues
\begin{equation}
E=\pm\left[
(v_{1x}q_x+v_{1y}q_y)^2+(v_{2x}q_x+v_{2y}q_y)^2
+(v_{3x}q_x+v_{3y}q_y)^2+(v_zq_z)^2
\right]^{1/2}.
\end{equation}
There is a Dirac point at $\bm{q}=0$.

Secondly, the effective Hamiltonian
invariant under $T$, $P$, and $\widetilde{M}^{\perp}_{x}$
has the form
\begin{equation}
H_0=q_x(v_{1x}\tau_x+v_{2x}\sigma_y\tau_z+v_{3x}\sigma_z\tau_z)
+(v_yq_y+v_zq_z)\sigma_x\tau_z,
\end{equation}
whose energy eigenvalues are
\begin{equation}
E=\pm\left[
(v_{1x}^2+v_{2x}^2+v_{3x}^2)q_x^2+(v_yq_y+v_zq_z)^2\right]^{1/2}.
\end{equation}
There is a Dirac line node located at $q_x=0$ and $v_yq_y+v_zq_z=0$.

Finally, 
the effective Hamiltonian invariant under
$T$, $P$, $\widetilde{C}^{\perp}_{2z}$,
and $\widetilde{M}^{\perp}_{x}$
has the form
\begin{equation}
H_0=q_x(v_{1x}\tau_x+v_{2x}\sigma_y\tau_z)+v_yq_y\sigma_x\tau_z,
\end{equation}
whose energy eigenvalues are
\begin{equation}
E=\pm\left[(v_{1x}^2+v_{2x}^2)q_x^2+(v_yq_y)^2\right]^{1/2}.
\end{equation}
There is a Dirac line located at $q_x=q_y=0$.
In all three cases, the influence of external magnetic field can be examined
by adding a Zeeman term
$\bm{H}\cdot\bm{\sigma}=h_x\sigma_x+h_y\sigma_y+h_z\sigma_z$ to $H_0$,
and the results of the analysis are summarized in Table I.

{\bf Single Dirac point/line node protected by off-centered screw/glide symmetries.}

\begin{figure*}[t]
\centering
\includegraphics[width=16 cm]{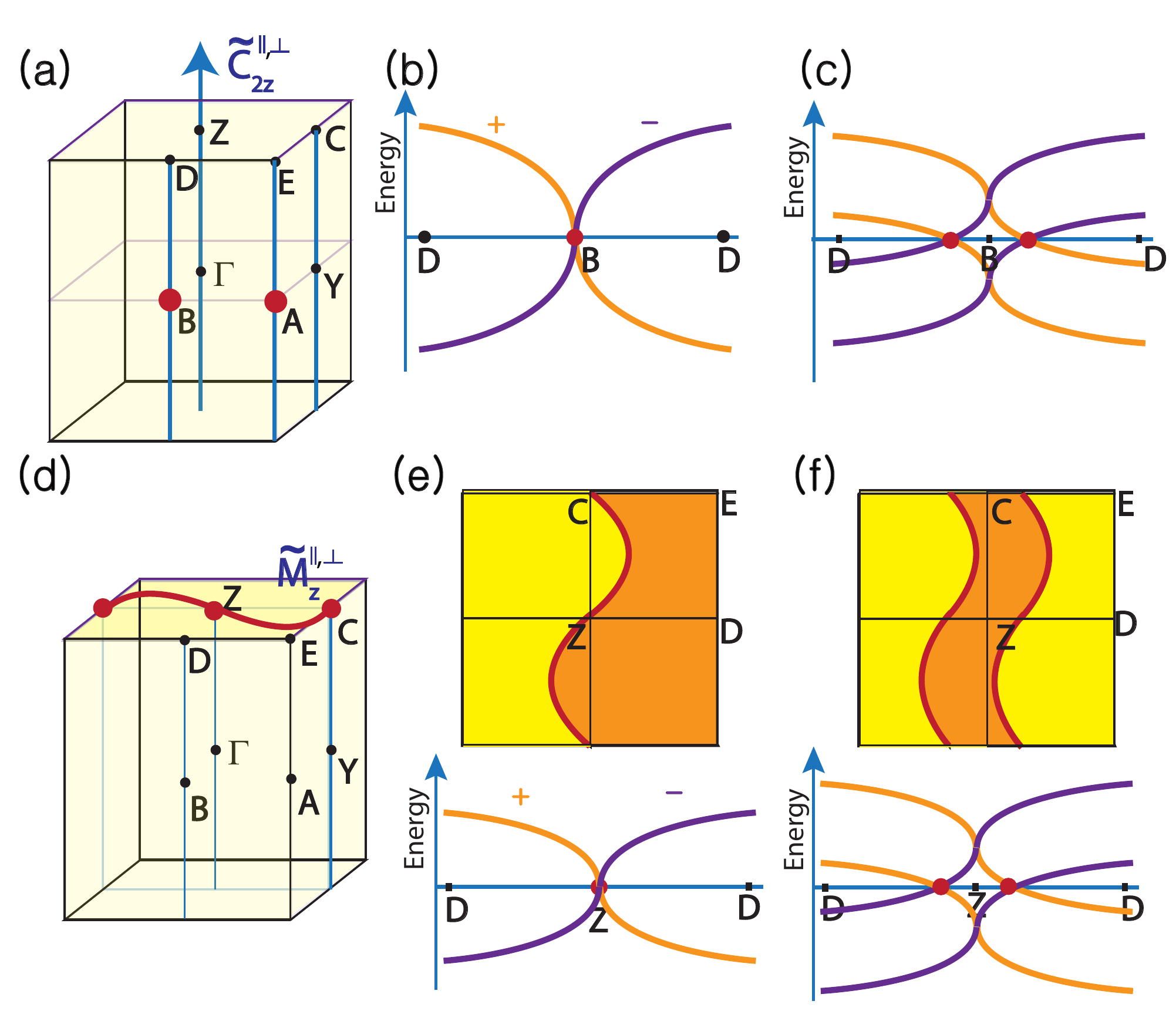}
\caption{
{\bf Dirac points/lines protected by off-centered screw/glide symmetries.}
({\bf a}-{\bf c}) Dirac points protected by off-centered two-fold
screw rotation $\widetilde{C}_{2z}^{\parallel,\perp}=\{C_{2z}|\frac{1}{2}\hat{x}+\frac{1}{2}\hat{z}\}$.
({\bf a})
A schematic figure describing the distribution of 3D Dirac points in momentum space.
The location of two 3D Dirac points is marked in red dots.
Here high symmetry momenta are labeled as in Ref.~\onlinecite{book}.
({\bf b})
The band structure along the $D-B-D$ line on which $\{P,\widetilde{C}_{2z}^{\parallel,\perp}\}=0$.
A pair of degenerate bands (a doublet pair) form a single 3D Dirac point at the time-reversal invariant momentum (TRIM) with $k_{z}=0$.
({\bf c})
The band structure when two doublet pairs cross on the $D-B-D$ line.
There are in total $2n$ ($n$ is an integer) 3D Dirac points, which are all away from TRIMs.
({\bf d}-{\bf f}) Dirac lines protected by off-centered glide
mirror $\widetilde{M}_{z}^{\parallel,\perp}=\{M_{z}|\frac{1}{2}\hat{x}+\frac{1}{2}\hat{z}\}$.
({\bf d})
A schematic figure describing the location of the nodal line in momentum space.
The location of the line node in the $k_{z}=\pi$ plane is marked in red color.
({\bf e})
Shape of a nodal line in the $k_{z}=\pi$ plane in which $\{P,\widetilde{M}_{z}^{\parallel,\perp}\}=0$
is satisfied.
The corresponding band structure along the  $D-Z-D$ line is shown in the bottom.
The doublet pair are degenerate at two TRIMs ($C$ and $Z$ points), which are a part of line nodes in the $k_{z}=\pi$ plane.
({\bf f})
Shape of nodal lines
when two doublet pairs cross in the $k_{z}=\pi$ plane.
The corresponding band structure along the $D-Z-D$ line is shown in the bottom.
} \label{fig:offcentered_glide}
\end{figure*}

Let us first consider an off-centered screw rotation $\widetilde{C}_{2z}^{\parallel,\perp}=\{C_{2z}|\frac{1}{2}\hat{x}+\frac{1}{2}\hat{z}\}$ which transforms
the spatial coordinate as
\begin{align}
\widetilde{C}_{2z}^{\parallel,\perp}:(x,y,z)\rightarrow (-x+\frac{1}{2},-y,z+\frac{1}{2}).
\end{align}
Let us note that the partial translation in $\widetilde{C}_{2z}^{\parallel,\perp}$ has both the parallel and perpendicular components relative to the rotation axis.
The eigenvalues of $\widetilde{C}_{2z}^{\parallel,\perp}$ are given by $c_{\pm}(k_{z})=\pm i e^{i\frac{1}{2}k_{z}}$
which is momentum dependent.
After straightforward calculation, one can find the following commutation relations
\begin{align}
\widetilde{C}_{2z}^{\parallel,\perp}PT&=e^{-ik_{x}+ik_{z}}PT\widetilde{C}_{2z}^{\parallel,\perp},
\nonumber\\
\widetilde{C}_{2z}^{\parallel,\perp}P&=e^{ik_{x}-ik_{z}}P\widetilde{C}_{2z}^{\parallel,\perp}.
\end{align}
Thus, along the $\widetilde{C}_{2z}^{\parallel,\perp}$ invariant lines with $k_{x}=\pi$, we find
\begin{align}
\widetilde{C}_{2z}^{\parallel,\perp}\left[PT|c_{\pm}(k_{z})\rangle\right]&=c_{\pm}(k_{z})\left[PT|c_{\pm}(k_{z})\rangle\right],
\end{align}
where $\widetilde{C}_{2z}^{\parallel,\perp}|c_{\pm}(k_{z})\rangle=c_{\pm}(k_{z})|c_{\pm}(k_{z})\rangle$.
Namely, the Kramers degenerate states related by $PT$ at a momentum $\bm{k}$ have the same 
$\widetilde{C}_{2z}^{\parallel,\perp}$ eigenvalues $c_{\pm}(k_{z})=\pm i e^{i\frac{1}{2}k_{z}}$.
From $c_{\pm}(k_{z}+2\pi)=c_{\mp}(k_{z})$, one can see that two sets of degenerate bands having different $\widetilde{C}_{2z}^{\parallel,\perp}$ eigenvalues should be interchanged when $k_{z}$ is shifted by $2\pi$. Also, the fact that $c_{\pm}(k_{z})$ is pure imaginary (real)
when $k_{z}=0$ ($k_{z}=\pi$) indicates that the two bands should be degenerate
at $k_{z}=0$ to satisfy time-reversal symmetry.
This consideration naturally leads
to the band structure shown in Fig.~\ref{fig:offcentered_glide} (a-c) where two sets of degenerate bands
having different $\widetilde{C}_{2z}^{\parallel,\perp}$ eigenvalues form a doublet pair with
a single Dirac-type crossing at $k_{z}=0$.
This doublet with a single Dirac point provides a basic building block
to construct the band structure along the $\widetilde{C}_{2z}^{\parallel,\perp}$ invariant lines with $k_{x}=\pi$.

Now we consider an off-centered glide mirror
$\widetilde{M}_{z}^{\parallel,\perp}=\{M_{z}|\frac{1}{2}\hat{x}+\frac{1}{2}\hat{z}\}$ which transforms
the spatial coordinate as
\begin{align}
\widetilde{M}_{z}^{\parallel,\perp}:(x,y,z)\rightarrow (x+\frac{1}{2},y,-z+\frac{1}{2}).
\end{align}
Let us note that the partial translation in $\widetilde{M}_{z}^{\parallel,\perp}$ has both the parallel and perpendicular components relative to the mirror plane.
The eigenvalues of $\widetilde{M}_{z}^{\parallel,\perp}$ are given by $m_{\pm}(k_{x},k_{y})=\pm i e^{i\frac{1}{2}k_{x}}$
which is momentum dependent.
After straightforward calculation, one can find the following commutation relations
\begin{align}
\widetilde{M}_{z}^{\parallel,\perp}PT&=e^{-ik_{z}+ik_{x}}PT\widetilde{M}_{z}^{\parallel,\perp},
\nonumber\\
\widetilde{M}_{z}^{\parallel,\perp}P&=e^{ik_{z}-ik_{x}}P\widetilde{M}_{z}^{\parallel,\perp}.
\end{align}
Thus, in the $k_{z}=\pi$ plane, we find
\begin{align}
\widetilde{M}_{z}^{\parallel,\perp}\left[PT|m_{\pm}(k_{x},k_{y})\rangle\right]&=m_{\pm}(k_{x},k_{y})\left[PT|m_{\pm}(k_{x},k_{y})\rangle\right],
\end{align}
where $\widetilde{M}_{z}^{\parallel,\perp}|m_{\pm}(k_{x},k_{y})\rangle=m_{\pm}(k_{x},k_{y})|m_{\pm}(k_{x},k_{y})\rangle$.
Namely, the degenerate states related by $PT$ at a momentum $\bm{k}$ have the same 
$\widetilde{M}_{z}^{\parallel,\perp}$ eigenvalues $m_{\pm}(k_{x},k_{y})=\pm i e^{i\frac{1}{2}k_{x}}$.
From $m_{\pm}(k_{x}+2\pi,k_{y})=m_{\mp}(k_{x},k_{y})$, one can see that two sets of degenerate bands having different $\widetilde{M}_{z}^{\parallel,\perp}$ eigenvalues should be interchanged when $k_{x}$ is shifted by $2\pi$. Also, the fact that $m_{\pm}(k_{x},k_{y})$ is pure imaginary 
when $(k_{x},k_{y})=(0,0)$, $(0,\pi)$ indicates that the two bands should be degenerate
at these two points to satisfy time-reversal symmetry.
This consideration naturally leads
to the band structure shown in Fig.~\ref{fig:offcentered_glide} (d-f) where two sets of degenerate bands
having different $\widetilde{M}_{z}^{\parallel,\perp}$ eigenvalues form a doublet pair in the whole
$k_{z}=\pi$ plane with
a single open-shaped Dirac line node passing $(k_{x},k_{y})=(0,0)$ and $(0,\pi)$.
This doublet with a single Dirac line node provides a basic building block
to construct the band structure in the $k_{z}=\pi$ plane.

\bibliographystyle{naturemag}

{\small \subsection*{Acknowledgements}
We are grateful to Robert-Jan Slager for useful comments.
B.-J. Y was supported by IBS-R009-D1, Research Resettlement Fund for the new faculty of Seoul National University, and Basic Science Research Program through the National Research Foundation of Korea (NRF) funded by the Ministry of Education (Grant No. 0426-20150011).
TM was supported by the EPiQS initiative of the Gordon and Betty Moore Foundation.
AF was supported by JSPS KAKENHI Grant (No.~15K05141).}

{\small \subsection*{Author contributions}
B.-J.Y., T.M. and A.F. conceived the original ideas.
T.A.B. carried out the numerical calculations.
B.-J.Y. and A.F. wrote the manuscript.
All authors discussed the results and commented on the manuscript.
}

{\small \subsection*{Additional information}
Reprints and permission information is available online at http://npg.nature.com/reprintsandpermissions/

Correspondence and requests for materials should be addressed to
B.-J.Y.\ or A.F..}

{\small \subsection*{Competing financial interests} The authors declare no competing financial interests.}


\newpage

\begin{table*}[h]
\begin{tabular}{c c c c c c c c c c c c c c c c}
\hline
\hline
Magnetic field ($\bm{H}$) & & $\widetilde{C}^{\perp}_{2z}$& & $\widetilde{M}^{\perp}_{x}$& & $\widetilde{M}^{\perp}_{y}$& &$T$ & &$P$ &&Fermi surface topology && Figures \\
\hline
\hline
$\bm{H}=0$ & & No & &Yes & & No & & Yes & & Yes & & Dirac line node && 4, 7(a)\\
$\bm{H}//[100]$ & & No & & Yes & & No & & No & & Yes & & Weyl line nodes && 7(b)\\
$\bm{H}//[010]$ & & No & & No & & No & & No & & Yes & & Gapped & \\
$\bm{H}//[001]$ & & No & & No & & No & & No & & Yes & & Gapped && 7(c)\\
\hline
\hline
$\bm{H}=0$ & & Yes & &No & & No & & Yes & & Yes & & Dirac point nodes && 5, 6(a)\\
$\bm{H}//[100]$ & & No & & No & & No & & No & & Yes & & Weyl point nodes &\\
$\bm{H}//[010]$ & & No & & No & & No & & No & & Yes & & Weyl point nodes & \\
$\bm{H}//[001]$ & & Yes & & No & & No & & No & & Yes & & Weyl point nodes && 6(b)\\
\hline
\hline
$\bm{H}=0$ & & Yes & &Yes & & Yes & & Yes & & Yes & & Dirac line nodes && 8(a-c)\\
$\bm{H}//[100]$ & & No & & Yes & & No & & No & & Yes & & Weyl line nodes && 8(d-f)\\
$\bm{H}//[010]$ & & No & & No & & Yes & & No & & Yes & & Weyl line nodes & \\
$\bm{H}//[001]$ & & Yes & & No & & No & & No & & Yes & & Gapped & \\
\hline
\hline
\end{tabular}
\end{table*}
{\noindent {\bf Table 1: Fermi surface topology of the semimetals protected by off-centered symmetries
in the presence/absence of magnetic field.}
{
Classification table for 3D topological semimetals
with nodal point/lines protected by off-centered symmetries
$\widetilde{C}^{\perp}_{2z}$, $\widetilde{M}^{\perp}_{x}$, $\widetilde{M}^{\perp}_{y}$
as well as time-reversal $T$ and the inversion $P$ symmetries
in the presence/absence of external Zeeman fields.
The ``presence" or ``absence" of each symmetry is indicated by ``Yes"
or ``No", respectively, in the table.
We note that the three off-centered symmetries are not independent
due to the relation $\widetilde{C}^{\perp}_{2z}=\widetilde{M}^{\perp}_{y}\widetilde{M}^{\perp}_{x}$.
Here a Dirac (Weyl) point/line node indicates a point/line node with four-fold (two-fold) degeneracy.
}
}

\newpage

\begin{table*}[h]
\begin{tabular}{c c c c c c c c c}
\hline
\hline
Candidate system & & Space group & & Relevant symmetry & & Shape of line nodes && Number of line nodes \\
\hline
\hline
$\hat{H}^{(0)}$ & & 59 & & $\widetilde{M}_{x}^{\perp}$ or $\widetilde{M}_{y}^{\perp}$ & & open straight && 2 \\  
$\hat{H}^{(0)}+\delta H^{(1)}(\bm{k})$ & & 11 & & $\widetilde{M}_{x}^{\perp}$ & & open && 1 \\  
BaTaS & & 194 & & $\widetilde{M}_{z}^{\perp}$ & & open straight && 3 \\  
SrIrO$_{3}$ && 62 && $\widetilde{C}_{2z}^{\parallel,\perp}$ and $\widetilde{C}_{2y}^{\parallel}$ && closed loop && 1 \\
\hline
\hline
\end{tabular}
\end{table*}
{\noindent {\bf Table 2: Properties of candidate systems
having Dirac line nodes with four-fold degeneracy
in the presence of strong spin-orbit coupling.}
{
The shape/number of Dirac line nodes, and the associated
symmetries in
two candidate materials BaTaS~\cite{linenode_new_Weng}, SrIrO$_3$~\cite{linenode_new_HY}, and the model Hamiltonian $\hat{H}^{(0)}$ in Eq.~(\ref{eqn:model_Rspace})
with/without the distortion described by $\hat{H}^{(1)}$ in Eq.~(\ref{eqn:Hpert1}).
These are the only examples proposed up to now which can support stable Dirac line nodes
with four-fold degeneracy even in the presence of strong spin-orbit coupling.
Here the number of line nodes indicates the minimal number of Dirac line nodes appearing
near the Fermi level in each system when the nodal lines are assumed
to be almost dispersionless.
}
}

\newpage

\end{document}